\documentclass[prb,twocolumn,balancelastpage,superscriptaddress,showpacs,floatfix,letterpaper]{revtex4-1}
\usepackage{times}
\usepackage{verbatim}
\usepackage{bm}
\usepackage{makeidx}
\usepackage{pifont}
\usepackage{amsmath}
\usepackage{amsfonts}
\usepackage{amscd}
\usepackage{amsthm}
\usepackage{amssymb}
\usepackage{graphicx}
\usepackage{hyperref}
\graphicspath{{./figures/}{/}}

\newcommand{\ddau}[2]                   {\frac{\partial #1}{\partial #2}}    
\newcommand{\ee}                        {e}                                  
\newcommand{\idmat}                     {\openone}                           
\newcommand{\ii}                        {i}                                  
\renewcommand{\vec}[1]                  {\mathbf{#1}}                        
\newcommand{\unitvec}[1]                {\vec{e}_{#1}}                       
\newcommand{\avg}[1]                    {\langle#1\rangle}                   
\newcommand{\lB}                        {l_B}                                

\newcommand{\tfracs}[2]               {{#1}/{#2}}                          
\newcommand{\tfracsb}[2]              {({#1}/{#2})}                        
\newcommand{\tfdd}[2]                 {{\mathrm{d} #1}/{\mathrm{d} #2}}        

\providecommand{\eqn}                   {Eq.~}

\providecommand{\sect}                  {Sec.~}
\providecommand{\sects}                 {Secs.~}

\newcommand{\coloronline} {(Color online)\ }

\providecommand{\abs}[1]                {\lvert#1\rvert}

\newcommand    {\vk}         {\vec{k}}
\newcommand    {\tR}         {t_\mathrm{R}} 
\newcommand    {\tI}         {t_\mathrm{I}} 
\newcommand    {\lambdaZ}    {\lambda_\mathrm{Z}} 
\newcommand    {\gex}        {g_\mathrm{ex}}
\newcommand    {\muB}        {\mu_\mathrm{B}} 
\newcommand    {\sigmaH}     {\sigma_\mathrm{H}} 
\newcommand    {\sigmaspH}   {\sigma_\mathrm{H}^{\mathrm{sp}}} 

\newcommand    {\subA}       {\mathrm{A}}  
\newcommand    {\subB}       {\mathrm{B}}  
\newcommand    {\matD}       {\mathcal{D}}  
\newcommand    {\matR}       {\mathcal{R}}  

\newcommand    {\spinup}     {\mathord{\uparrow}}
\newcommand    {\spindown}   {\mathord{\downarrow}}

\newcommand    {\nn}[1]      {\left\langle #1 \right\rangle}
\newcommand    {\nnn}[1]     {\left\langle\langle #1 \right\rangle\rangle}
\newcommand    {\vecsigma}   {\bm{\sigma}}

\begin{document}
\title{Topological phases in a two-dimensional lattice: Magnetic field versus spin-orbit coupling}
\author{W. Beugeling}
\affiliation{Institute for Theoretical Physics, Utrecht University, Leuvenlaan 4, 3584 CE Utrecht, The Netherlands}
\author{N. Goldman}
\affiliation{Center for Nonlinear Phenomena and Complex Systems, Universit\'e Libre de Bruxelles (U.L.B.), B-1050 Brussels, Belgium}
\author{C. \surname{Morais Smith}}
\affiliation{Institute for Theoretical Physics, Utrecht University, Leuvenlaan 4, 3584 CE Utrecht, The Netherlands}
\date{August 13, 2012}

\begin{abstract}
In this work, we explore the rich variety of two-dimensional topological phases that arise when considering the competing effects of spin-orbit couplings, Zeeman splitting and uniform magnetic fields. We investigate minimal models, defined on a honeycomb lattice, which clarify the topological phases stemming from the intrinsic and Rashba spin-orbit couplings, and also from the Zeeman splitting. In this sense, our work provides an interesting path connecting the quantum Hall phases, generally produced by the uniform magnetic field, and the quantum spin Hall phases resulting from spin-dependent couplings. First, we analyze the properties of each coupling term individually and we point out their similarities and differences. Secondly, we investigate the subtle competitions that arise when these effects are combined. We finally explore the various possible experimental realizations of our model.
\end{abstract}

\maketitle

\section{Introduction}

Today, unprecedented efforts are devoted to the study of novel topological insulating phases, which exhibit remarkable properties at their boundaries.\cite{HasanKane2010,QiZhang2011} In two dimensions, these quantum states are characterized by propagating edge states, which carry dissipationless currents along one-dimensional boundaries. From a theoretical point of view, these transport properties rely on the existence of large bulk gaps in the energy spectrum, which host robust and gapless edge excitations. Different edge state structures and transport properties can be realized according to the nature of the gaps: A magnetic field breaks time-reversal symmetry (TRS) and leads to chiral edge states (i.e., particles propagating in a given direction) and quantized Hall currents,\cite{VonKlitzingEA1980} whereas a large spin-orbit coupling preserves the TRS and produces helical edge structures.\cite{KaneMele2005PRL95-14,KaneMele2005PRL95-22} In the latter situation, referred to as the quantum spin Hall (QSH) effect, edge states with opposite spins counterpropagate and contribute to a vanishing Hall current, while producing a spin Hall current.\cite{KonigEA2007,BernevigEA2006} The origins of the quantum Hall (QH) and QSH phases are fundamentally different, as the former is produced by an external field, whereas the latter relies on the intrinsic properties of the material.  However, one can interpret the QSH phase as being two opposite QH phases, one for each spin component. This observation is easily shown through the Kane-Mele lattice model,\cite{KaneMele2005PRL95-14,KaneMele2005PRL95-22} which realizes the QSH effect by the inclusion of the intrinsic spin-orbit coupling (ISO), and corresponds to two copies of the (QH) Haldane model.\cite{Haldane1988}

\par  In recent works, the ISO coupling has been combined with a magnetic field,\cite{GoldmanEA2012,ShevtsovEA2012} or with a constant exchange term,\cite{YangEA2011PRL} in order to study the effect of the TRS breaking on the QSH effect. It has been shown that the QSH effect persists in the presence of magnetic field, leading to the ``TRS-broken''\cite{YangEA2011PRL} or ``weak'' QSH effect.\cite{GoldmanEA2012} The term ``weak'' refers to the absence of robustness against spin-flip scattering due to magnetic disorder. In absence of this scattering process, the Hall and spin Hall conductivities are protected and quantized.

\par In this work, we present a different approach to study the interplay between QH and QSH physics. Contrarily to the general trend in the field of topological insulators, which is to include many terms and study complicated Hamiltonians, here we investigate the minimal models that can produce topological phases. We consider a tight-binding model of the honeycomb lattice under a perpendicular magnetic field, which includes the ISO coupling, the Rashba spin-orbit (RSO) coupling, and the Zeeman effect. This spin-1/2 model was introduced in our recent paper,\cite{GoldmanEA2012} where we already studied the combined effects of some of these terms. Here, instead, we first investigate the effect of each of these terms individually on the QH phases generated by the magnetic field, in order to distill the problem. In particular, we demonstrate the equivalence of the ISO and Zeeman terms in generating the weak QSH phase at zero energy in the presence of a magnetic field. Then, we show that the ISO coupling can drive topological phase transitions between different topological insulating phases, an effect which totally relies on the combination of the ISO coupling and the magnetic field.

\par Secondly, we study the interplay between the terms incorporated in our model. For example, we discuss the competition between the ISO coupling and Zeeman effect, motivated by the similarity of these terms for generating the weak QSH effect. We show that the combination of these terms does not necessarily lead to a larger regime of parameters where the weak QSH effect is observed, which reveals the subtle competition between each of these terms and the magnetic field. We also study the exquisite effects produced by the Rashba coupling in the presence of a large exchange field, which is shown to generate several QH gaps, even in the absence of a magnetic field. From another perspective, we also discuss the fate of the weak QSH phase, as the RSO coupling is turned on, which introduces spin-flip terms into the Hamiltonian, so that the spins are no longer aligned perpendicularly to the plane, and the spin Hall conductivity is no longer protected.

\par An important motivation for the investigation of these topological phases and phase transitions is the rich variety of systems where they may be observed. Despite its weak ISO coupling, graphene remains an appealing candidate for the observation of these phases, e.g., by increasing the ISO coupling artificially by adatoms.\cite{WeeksEA2011} Secondly, several experimental groups have reported on synthetic honeycomb lattices in a condensed matter system, for instance arrays of quantum dots on a GaAs substrate,\cite{SinghaEA2011} and ``molecular graphene'', where the lattice is created by using the repulsive interactions of CO molecules deposited in a triangular array on a Cu(111) surface.\cite{GomesEA2012} These systems have larger lattice constants, which lead to a higher flux per plaquette at realistic magnetic field values, and allow for a more flexible control over the parameters than in real graphene. A third type of experiments that motivates our studies is the recent realization of a honeycomb optical lattice for ultracold atoms.\cite{SoltanPanahiEA2011,TarruellEA2012,LeeEA2009} In cold-atom systems,  large magnetic fields\cite{LinEA2009} and spin-orbit couplings\cite{LinEA2011Nature} are produced synthetically by adjustable lasers (see Ref.~\onlinecite{DalibardEA2011} for a review). In particular, such artificial gauge potentials could be produced in optical lattices,\cite{DalibardEA2011,AidelsburgerEA2011,Cooper2011,GoldmanEA2010,*MazzaEA2012,*OsterlohEA2005,*BeriCooper2011} thus leading to the possibility to probe the high-flux regime. Furthermore, we note that optical \emph{square} lattices subjected to well-designed gauge potentials can reproduce the properties of honeycomb lattice systems, exhibiting  Dirac-type physics.\cite{LimEA2008,*LimEA2010,GoldmanEA2009,BermudezEA2010PRL,LanEA2011} Cold-atoms systems carry the advantages of high tunability, necessary for studying phase transitions, and allow for the study of topological phases in absence of interparticle interactions \cite{ChinEA2010} and disorder.\cite{BlochEA2008} These possible experimental realizations are explored in detail in the final part of this work.

This article is structured as follows. We introduce the lattice model in \sect\ref{sect_model} and present the four terms characterizing our tight-binding Hamiltonian: the usual hopping term, the two spin-orbit terms as well as the Zeeman splitting. Then, in \sect\ref{sect_tools}, we review the techniques used for our investigation and characterization of topological phases. Section~\ref{sect_magnetic_field} discusses the combined effect of the magnetic field and spin-orbit coupling. Then, the combined effects of spin-orbit couplings, Zeeman splitting and uniform magnetic flux are explored in \sect\ref{sect_competition_and_phasetrans}. The possible experimental realizations of our model, as well as the possibility to detect the effects presented in \sects\ref{sect_magnetic_field} and \ref{sect_competition_and_phasetrans}, are reported in \sect\ref{sect_experiment}. Finally, the conclusions are drawn in \sect\ref{sect_conclusion}. The Appendix provides further mathematical details of the calculations.

\section{Model}
\label{sect_model}

We consider a tight-binding model of spinful electrons in a two-dimensional honeycomb lattice, subjected to a uniform perpendicular magnetic field $\vec{B}=B\unitvec{z}$. This model was introduced in Ref.~\onlinecite{GoldmanEA2012} to investigate the effects of an external magnetic field on the Kane-Mele model (cf.\ Refs.~\onlinecite{KaneMele2005PRL95-14,KaneMele2005PRL95-22}). The magnetic field induces a Zeeman effect, and causes the hopping terms to acquire a phase. This phase is encoded by the so-called Peierls substitution, where one replaces the momentum $\vec{p}$ by $\vec{p}-e\vec{A}$, where $\vec{A}$ is the gauge potential associated with the magnetic field. Thus, any hopping term from site $k$ to site $j$ picks up the phase factor $\ee^{\ii\theta_{jk}}$, where $\theta_{jk}=(e/\hbar)\int_{\vec{r}_k}^{\vec{r}_j}\vec{A}\cdot\vec{d l}$. In the remainder of this text, we will express the magnetic field strength (flux density) in terms of the dimensionless quantity $\phi$, defined as the flux per unit cell of the lattice, expressed in units of the elementary flux quantum $h/e$.

\par In our model, we include four effects into the Hamiltonian, written as  $H=H_\mathrm{NN}+H_\mathrm{Z}+H_\mathrm{I}+H_\mathrm{R}$, with
\begin{align}\label{eqn_hamiltonian}
  H_\mathrm{NN}&=-t\sum_{\nn{j,k}}\ee^{\ii\theta_{jk}}c_j^\dagger c_k,\nonumber\\
  H_\mathrm{Z}&=-2\pi\phi\lambdaZ\sum_jc_j^\dagger \sigma_z c_j,\\
  H_\mathrm{I}&=-\ii \tI\sum_{\nnn{j,k}}\ee^{\ii\theta_{jk}}\nu_{jk}c_{j}^\dagger\sigma_z c_{k},\nonumber\\
  H_\mathrm{R}&=-\ii \tR\sum_{\nn{j,k}}\ee^{\ii\theta_{jk}}c_{j}^\dagger(\vecsigma \mathop{\wedge} \vec{d}_{jk})  c_{k}.\nonumber
\end{align}
Here, $c_k=(c_{k\,\spinup},c_{k\,\spindown})$ is the annihilation operator for electrons at site $k$.
The first term $H_\mathrm{NN}$ describes hopping between two nearest-neighbor (NN) sites $j$ and $k$, where $t$ denotes the hopping amplitude. The second term $H_\mathrm{Z}$ describes the Zeeman effect, which is an on-site term that assigns different potentials to the two spin components via the Pauli matrix $\sigma_z$. Its amplitude is proportional to the magnetic flux $\phi$ and the coefficient $\lambdaZ$, which is related to the material's Land\'e $g$ factor by $2\pi\phi\lambdaZ=g\muB B$, where $\muB$ is the Bohr magneton. Note that in a cold-atom emulation of this model, the parameters $\phi$ and $\lambdaZ$ are tuned individually (cf.\ \sect\ref{sect_experiment_optical_lattices}). The Zeeman term lifts the degeneracy by shifting the two spin-degenerate copies of the spectrum up and down by an equal amount of energy. The third term $H_\mathrm{I}$ describes the ISO coupling, which corresponds to a next-nearest-neighbor (NNN) hopping with amplitude $\tI$.\cite{KaneMele2005PRL95-14} Here, the factor $\nu_{jk}=\pm 1$, where the sign depends on the value of $\vec{d}_{kl}\mathop{\wedge}\vec{d}_{lj}$, i.e., the outer product of the two bond vectors connecting site $k$ to site $j$ via their unique common neighbor at site $l$. This hopping also involves the matrix $\sigma_z$, so that the sign of the hopping amplitude is opposite for the two spin components $\spinup$ and $\spindown$. Thus, the two different spin species are effectively subjected to opposite  local (Haldane-type \cite{Haldane1988}) magnetic flux. In absence of a magnetic field, this term opens a topologically nontrivial gap, and causes the QSH effect, characterized by so-called helical edge states, which are protected by TRS.\cite{KaneMele2005PRL95-14,KaneMele2005PRL95-22}
The final term $H_\mathrm{R}$ is the contribution to the NN hopping due to the RSO coupling. This hopping has an amplitude $\tR$, and involves the spin matrix $\vecsigma\mathop{\wedge} \vec{d}_{jk}=\sigma_x d^y_{jk}-\sigma_y d^x_{jk}$, where $\vecsigma=(\sigma_x,\sigma_y)$, and $\vec{d}_{jk}$ is the vector connecting sites $k$ and $j$. The resulting spin matrix has only off-diagonal elements, so that this hopping involves a spin flip. This term couples the two spin components, and as a consequence, $\sigma_z$ is no longer a good quantum number for $\tR \ne 0$. In practice, this means that spin states are no longer exclusively up or down, but may also point in different directions. Finally, we remark that the model considered here preserves inversion symmetry. In particular, we disregard the effect of an additional staggered potential, which was shown to induce topological phase transitions between trivial and non-trivial phases in Refs.~\onlinecite{KaneMele2005PRL95-14,KaneMele2005PRL95-22}. The possible physical realizations of our model \eqref{eqn_hamiltonian} are discussed later in \sect\ref{sect_experiment}.

\section{Tools for the topological analysis}%
\label{sect_tools}%
In the following sections, we will perform an analysis of the various topological phases produced by the terms in the model Hamiltonian~\eqref{eqn_hamiltonian}. Before going into details, describing each term individually and their combined effects, we first present the framework and tools which we will use for this analysis. Readers who are already familiar with the notions of bulk and edge states, Chern numbers, and (spin) Hall conductivity may choose to skip this section.

\subsection{Harper equation and the Hofstadter butterfly}%
\label{sect_harper_hofstadter}%
We will illustrate our framework by studying the spin-degenerate model, i.e., we set $\lambdaZ=\tI=\tR=0$ in Hamiltonian \eqref{eqn_hamiltonian}. We assume $t>0$ throughout the whole text. The bulk band structure can be computed numerically by applying periodic boundary conditions, namely by considering a toroidal geometry. This requires to set $\phi=p/q$, where $p$ and $q$ are integers, in which case the system is periodic in both spatial directions.  Under these conditions, and by choosing a proper gauge for the Peierls phases $\ee^{\ii\theta_{jk}}$, the system reduces to a $q\times 1$ magnetic unit cell. (See the Appendix for a description of the gauge structure and its spatial periodicity.) By applying a Fourier transform and invoking Bloch's theorem, the Schr\"odinger equation reduces to a $4q\times4q$ eigenvalue problem, known as the discrete Harper equation,\cite{Hofstadter1976,[{In a more general form, the Harper equation is known as an \emph{almost Mathieu equation} in mathematical literature; see e.g., }][{}]{BellissardSimon1982}}
\begin{equation}
(E/t)\Psi_n=\matD_n\Psi_n+\matR_n\Psi_{n+1}+\matR^\dagger_{n-1}\Psi_{n-1},\label{eqn_harper_main}
\end{equation}
where $\Psi_{n}=(\psi_{n\,\subA\,\spinup},\psi_{n\,\subA\,\spindown},\psi_{n\,\subB\,\spinup},\psi_{n\,\subB\,\spindown})$ denotes the single-particle wave function at site $n=1, \dots ,q$,  and where the $4 \times 4$ matrices $\matD_n$ and $\matR_n$ are given in terms of the parameters defined in Hamiltonian \eqref{eqn_hamiltonian} (see the Appendix for details). The four components of the wave function  $\Psi_{n}$ are due to the two sublattice ($A$,$B$) and the two spin ($\uparrow$,$\downarrow$) degrees of freedom. The $4q\times4q$ Harper problem in \eqn\eqref{eqn_harper_main} is analogous to the original (spinless, square-lattice) Hofstadter problem,\cite{Hofstadter1976} where the Harper equation involves a $q\times q$ matrix.

\begin{figure}
\includegraphics{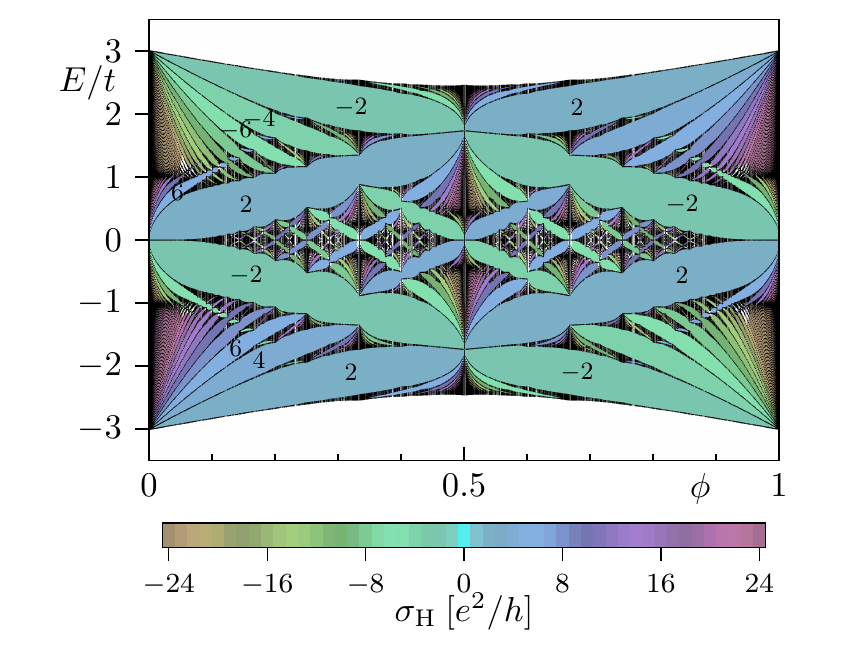}
\caption{\label{fig_spectrum_spinless}\coloronline Hofstadter butterfly spectrum for the spin-degenerate model, Hamiltonian \eqref{eqn_hamiltonian} with $\lambdaZ=\tI=\tR=0$. The bulk bands, i.e., the regions where the density of bulk states is positive, are shown in black. In the bulk gaps, the colors and numbers indicate the quantized Hall conductivity in units of $e^2/h$.}
\end{figure}

\par Solving the Harper equation \eqref{eqn_harper_main} provides the bulk energy bands $E=E(k_x,k_y)$. It is instructive to evaluate the sizes of the bulk gaps as a function of the magnetic flux $\phi$, which results in a diagram known as the  Rammal-Hofstadter butterfly,\cite{Hofstadter1976,Rammal1985} shown in Fig.~\ref{fig_spectrum_spinless}. This figure may be obtained in the limit of large $q$, where the bands become dispersionless. The fractal structure, which is periodic in $\phi$ with a period of $1$, is a result of the competition between two length scales: The magnetic length $\lB=\sqrt{\hbar/eB}$ and the lattice spacing. Here, $\phi$ parametrizes this competition, as the ratio between the area of the unit cell and that of the cyclotron orbits ($\sim\lB^2$).

\begin{figure}
\includegraphics{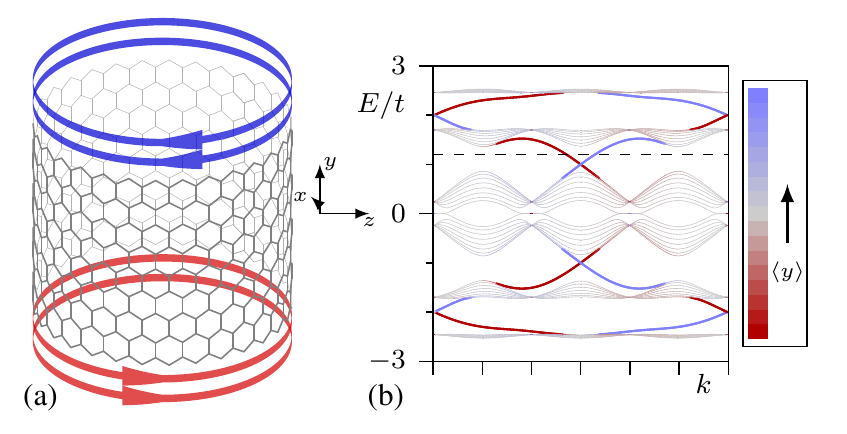}
\caption{\label{fig_cylinder_edge_states}\coloronline (a) The cylinder geometry representing a ribbon with periodic boundary conditions in the $x$ direction. (b) Spectrum at $\phi=1/3$ for the spin degenerate model ($\lambdaZ=\tI=\tR=0$). The colors represent the expectation value $\avg{y}$: Red and blue indicate edge states at the bottom and top edge, respectively, and gray represents bulk states. The  spin-degenerate edge states shown in (a) correspond to those in the QH gap at the Fermi energy indicated by the dashed line in (b).}
\end{figure}

\subsection{Topological invariants}

\par Each bulk gap in the spectrum shown in Fig.~\ref{fig_spectrum_spinless} may be characterized by a topological invariant, which encodes the Hall conductivity of this system when the Fermi energy lies inside this gap. The  integer values of the Hall conductivity are robust against external perturbations: They remain constant as long as the bulk gaps remain open. In the following, we recall the relation between this transport coefficient and the concept of edge states and topological invariants.

\subsubsection{Edge-state analysis}

The edge-state analysis can be performed for a system featuring a boundary,\cite{Hatsugai1993PRB} such as a cylinder, as shown in Fig.~\ref{fig_cylinder_edge_states}(a). When solving the Harper equation \eqref{eqn_harper_main} on a cylinder, namely by applying periodic boundary conditions along one spatial direction only, the spectrum consists of \emph{bulk bands} and \emph{topological edge states}.\cite{Hatsugai1993PRB} Indeed, we typically find a few \emph{edge states} within the bulk gaps, some of which \emph{cross} the gap from one bulk band to the other; see Fig.~\ref{fig_cylinder_edge_states}(b). Importantly, each edge state contributes $e^2/h$ (one quantum of conductivity) to the Hall conductivity of the system. The total Hall conductivity is therefore equal to an integer number (i.e., the number of edge-state branches) times $e^2/h$. The topological invariance of the Hall conductivity is due to the fact that the number of edge states inside the bulk gap cannot be altered unless the gap is closed. This result can be used to derive the topological invariant associated to any of the bulk gaps, leading to the colors in the spectrum in Fig.~\ref{fig_spectrum_spinless}. For this spin-degenerate model, all topological invariants are multiples of $2$ because the system is twofold spin degenerate.

\par More precisely, in order to evaluate the Hall conductivity of a bulk gap, we count the edge states whose dispersions intersect the Fermi energy, taking into account their location and direction of propagation. The location of each state is derived by computing its expectation value $\avg{y}$ from the eigenstate, which provides the coloring of Fig.~\ref{fig_cylinder_edge_states}(b). Secondly, the direction of propagation can be obtained from the sign of its momentum derivative $\tfdd{E}{k}$, where $k$ is the momentum parallel to the edge. Edge states with opposite directions contribute with opposite signs to the Hall current and therefore to the Hall conductivity,
\begin{equation}\label{eqn_sigmaH_right_left}
  \sigmaH=(N_\mathrm{R}-N_\mathrm{L})\frac{e^2}{h},
\end{equation}
where $N_\mathrm{L}$ and $N_\mathrm{R}$ denote the number of left- and right-moving states, respectively. The Hall conductivity does not depend on the magnitude of the velocity ($\propto\tfdd{E}{k}$), but instead is given by the number of edge-state channels.\cite{ButtikerEA1985} In this spin-degenerate model, the states are all either left- or right-moving, i.e., they are \emph{chiral} states. The chirality of the edge modes provides protection of the Hall conductivity against backscattering due to disorder in the presence of broken TRS.

\subsubsection{Bulk topological invariants}

\par Remarkably, the number of edge states is directly related to topological invariants, the \emph{Chern numbers}, a quantity associated with each of the bulk bands. The Chern number is an integer topological index  defined in the  toroidal geometry,\cite{Kohmoto1985,Hatsugai1993PRL}
\begin{align}\label{eqn_chernnumber}
  C_n
  =\frac{1}{2\pi\ii}\int_{\mathrm{BZ}}\left(\ddau{\vec{A}_{n,\vk}^y}{k_x}-\ddau{\vec{A}_{n,\vk}^x}{k_y}\right)d\vk,
\end{align}
where $\vec{A}_{n,\vk}= \langle u_{n,\vk}|\boldsymbol{ \nabla}_\vk|u_{n,\vk}\rangle$ is the Berry connection associated to the eigenstate $|u_{n,\vk}\rangle$, and where BZ denotes the (toroidal) Brillouin zone. The number of edge states $N_r$  inside the $r$th bulk gap, or equivalently the Hall conductivity assigned to this gap (cf.\ discussion above), is then equal to the sum of all the Chern numbers associated to the occupied bands,\cite{ThoulessEA1982,HasanKane2010}
\begin{equation}\label{eqn_bbc}
  N_r=\sum_{n\in{}\text{occupied bands}}C_{n}.
\end{equation}
This identity is known as the \emph{bulk-boundary correspondence},\cite{Hatsugai1993PRB,Hatsugai1993PRL} as it relates the topological indices associated to the bulk bands to the number of edge excitations. This correspondence provides us with an important observation: Although the dispersions of the edge states depend on the shape of the edges (e.g., zigzag, bearded, or armchair) and on the system size, the number of edge-state branches does not: this number is a bulk property and therefore does not depend on the specific form of the edge. In the remainder of this paper, we will study systems with zigzag and bearded edges, without any loss of generality concerning the topological properties.

\par In the next sections, we will characterize the topological phases by their associated Hall conductivity and edge-state structures. As described in the following, the edge states produced by the magnetic field are affected in the presence of strong perturbing terms, such as the spin orbit couplings and Zeeman splitting. For the sake of simplicity, we will show these effects using the edge-state analysis obtained from a cylindrical geometry, rather than focusing on the analysis of the bulk topological invariants.

\subsubsection{The St\v{r}eda formula}

\par An efficient method to compute the Hall conductivity can be formulated in terms of the integrated density of states. Inside a bulk gap, the integrated density of states $N_\phi(E)$ is defined as the fraction of all states that lie below the gap (which is an integer times $1/4q$). The Hall conductivity is then proportional to the $\phi$-derivative of this quantity,
\begin{equation}\label{eqn_streda_formula}
  \sigma_\mathrm{H}=4\frac{e^2}{h}\ddau{N}{\phi}.
\end{equation}
a result known as the St\v{r}eda formula.\cite{Streda1982JPhysC15-22,*Streda1982JPhysC15-36,UmucalilarEA2008,BermudezEA2010NJP}  We have used this powerful formula to compute the Hall conductivity, and generate the colorful butterfly spectra presented in the remainder of this paper.

\subsubsection{Charge and spin Hall conductivities}

In the presence of terms involving the matrix $\sigma_z$ in the Hamiltonian, the spin $\mathrm{SU}(2)$ symmetry is broken. If spin-flip terms (i.e., linear combinations of $\sigma_x$ and $\sigma_y$) are absent, the spin in the $z$ direction is conserved, i.e., it is a good quantum number. In this case, which happens if the ISO coupling or the Zeeman splitting is present while the RSO is not, the Hamiltonian may be decomposed into the two spin components $(\uparrow, \downarrow)$. Thus, one is able to define the component-wise Chern numbers $C_{\spinup,n}$ and $C_{\spindown,n}$ and Hall conductivities $\sigma_{\mathrm{H}\,\spinup}$ and $\sigma_{\mathrm{H}\,\spindown}$. The charge Hall conductivity of the whole system is the sum of the two component-wise conductivities, $\sigmaH=\sigma_{\mathrm{H}\,\spinup}+\sigma_{\mathrm{H}\,\spindown}$, since both of them contribute equally to the Hall current. Considering their contributions to the spin current, spin-up and spin-down edge modes have opposite ``spin charges'' $\pm\hbar/2$. Thus, the spin Hall conductivity is equal to the difference of the two component-wise conductivities, $\sigmaspH/(e/4\pi)=(\sigma_{\mathrm{H}\,\spinup}-\sigma_{\mathrm{H}\,\spindown})/(e^2/h)$, where $e/4\pi$ is the elementary quantum of spin Hall conductivity.

\par In the spin-degenerate model discussed earlier, the two component-wise Hall conductivities are always equal, so that all spin Hall conductivities vanish. This is certainly different if the spin degeneracy is broken. For instance, in the presence of ISO, the QSH gap at zero magnetic field and at zero energy \cite{KaneMele2005PRL95-22} is characterized by $\sigma_{\mathrm{H}\,\spinup}=-\sigma_{\mathrm{H}\,\spindown}=-e^2/h$. Here, the contributions to the charge Hall conductivity cancel, but the contributions to the spin Hall conductivity add up to $\sigmaspH=-2e/4\pi$.

\par If the Hamiltonian contains spin-flip terms, e.g., the Rashba coupling, then the spin is not conserved, and the aforementioned definition of the spin Hall conductivity is no longer valid. However, the related spin Chern numbers, which equal $C^\mathrm{sp}_n=C_{\spinup,n}-C_{\spindown,n}$ if spin is conserved, remain well-defined topological invariants, even in absence of spin conservation.\cite{ShengEA2006,Prodan2009}

\section{Effects of the magnetic field}
\label{sect_magnetic_field}

\subsection{The spin-degenerate model: Landau Levels and the anomalous quantum Hall effect}
\label{sect_spindeg}

\par Zooming in on the butterfly of Fig.~\ref{fig_spectrum_spinless} near $\phi\approx 0$, we observe that the thickness of the bands decreases in the low-flux regime (not shown). Thus, in this limit we can treat the bands as being infinitely thin, i.e., as Landau levels. In the low-energy regime, between the Van Hove singularities at $E/t=\pm1$, the band structure at low flux coincides with the Landau-level spectrum known from graphene,\cite{CastroNetoEA2009,Goerbig2011RMP} with energies $E\propto \pm\sqrt{2\pi l \phi}$ (where $l=0,1,2,\ldots$ is the Landau-level index).

Moreover, computing the Hall conductivity in this region, we find the ``anomalous'' Hall-conductivity sequence $\sigma_H= 4(n+\tfracs{1}{2})e^2/h$ ($n\in\mathbb{Z}$), as observed in graphene.\cite{NovoselovEA2005,*ZhangEA2005} The breakdown of this structure around $E/t=\pm1$ can be understood from the large density of states near the Van Hove singularities and from the Chern numbers associated to these many bands. \cite{HatsugaiEA2006} The edge-state structures, Hall conductivity plateaus and topological aspects of spinless electrons in a honeycomb lattice subjected to a magnetic field have been thoroughly described in Ref.~\onlinecite{HatsugaiEA2006}.

\subsection{Intrinsic spin-orbit coupling}
\label{sect_iso}%

\begin{figure}
\includegraphics{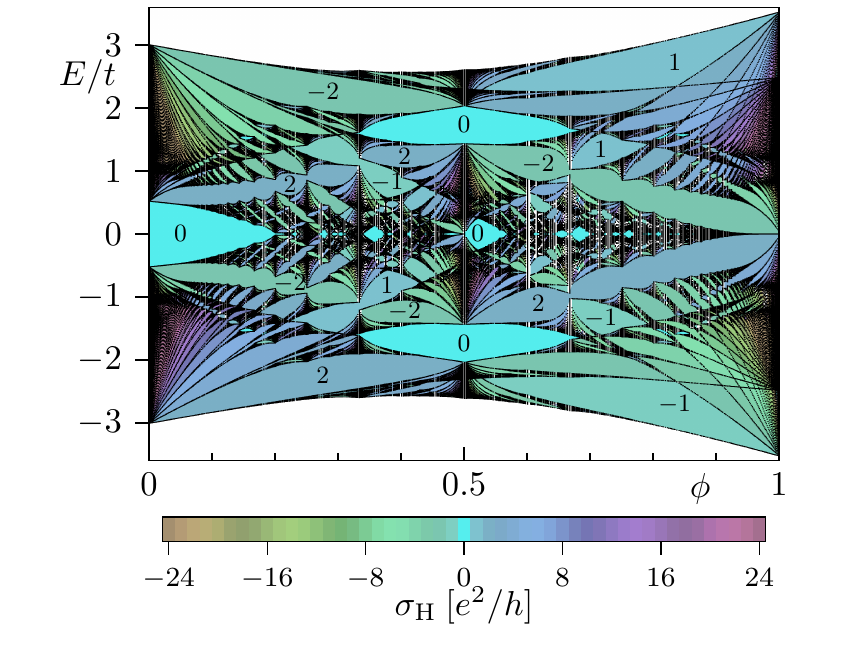}
\caption{\label{fig_spectrum_iso}\coloronline Hofstadter butterfly spectrum for the system with ISO coupling ($\tI/t=0.1$). Although the periodicity in $\phi$ is $6$ rather than $1$ (see the Appendix), we present only the range $0\leq \phi\leq 1$ for the sake of comparison with the other terms. The colors and numbers indicate the quantized Hall conductivity in units of $e^2/h$.}
\end{figure}

\par Since the seminal work of Kane and Mele,\cite{KaneMele2005PRL95-14} it is known that the ISO coupling opens a topologically nontrivial gap of size $\Delta=6\sqrt{3}\tI$ at zero magnetic field. This bulk gap hosts two counterpropagating edge modes per edge, with opposite spins (one Kramers pair): These \emph{helical} edge states are related by TRS. This topological phase is known as the QSH state, and may be regarded as two opposite QH phases (i.e., each spin performs the QH effect, with opposite chirality). The system preserves TRS (due to absence of the magnetic field), which protects the QSH state against scattering processes caused by disorder. Although for any edge state a reversely propagating mode would be available to scatter to, the spin has to be flipped, and the two different ways to flip the spin interfere destructively with each other, due to a mutual phase factor of $\ee^{\ii\pi}=-1$.\cite{QiZhang2010}

\par The combination of the ISO coupling and a magnetic field leads to  the breaking of the TRS  satisfied by the QSH phase. However, due to the absence of spin flip terms in the Hamiltonian, the helical edge-state structure persists. In terms of charge and spin Hall conductivity, the resulting state is equivalent to the QSH state. As announced in the introduction, this state is called the \emph{weak QSH} phase\cite{GoldmanEA2012} or \emph{TRS broken QSH} phase.\cite{YangEA2011PRL} From the Hofstadter butterfly in Fig.~\ref{fig_spectrum_iso}, we observe that the size of the weak QSH gap at zero energy tends to decrease if the flux is increased (but not monotonically), and closes eventually.  In Fig.~\ref{fig_spectrum_iso}, in which $\tI=0.1t$, the QSH gap at $E=0$ remains open in the range $0\leq\phi\leq0.2$. Here we observe the competition between the ISO coupling, which opens the (weak) QSH gap, and the magnetic field, which ``tries'' to destroy the weak QSH state by its TRS breaking property. Throughout the weak QSH gap, the edge states remain crossing at zero energy, which indicates that this phase is robust at least in the absence of magnetic disorder. We finally note that the spectrum remains particle-hole symmetric, as in the spinless case.

\subsection{Zeeman effect}
\label{sect_zeeman_flux}%

\begin{figure}
\includegraphics{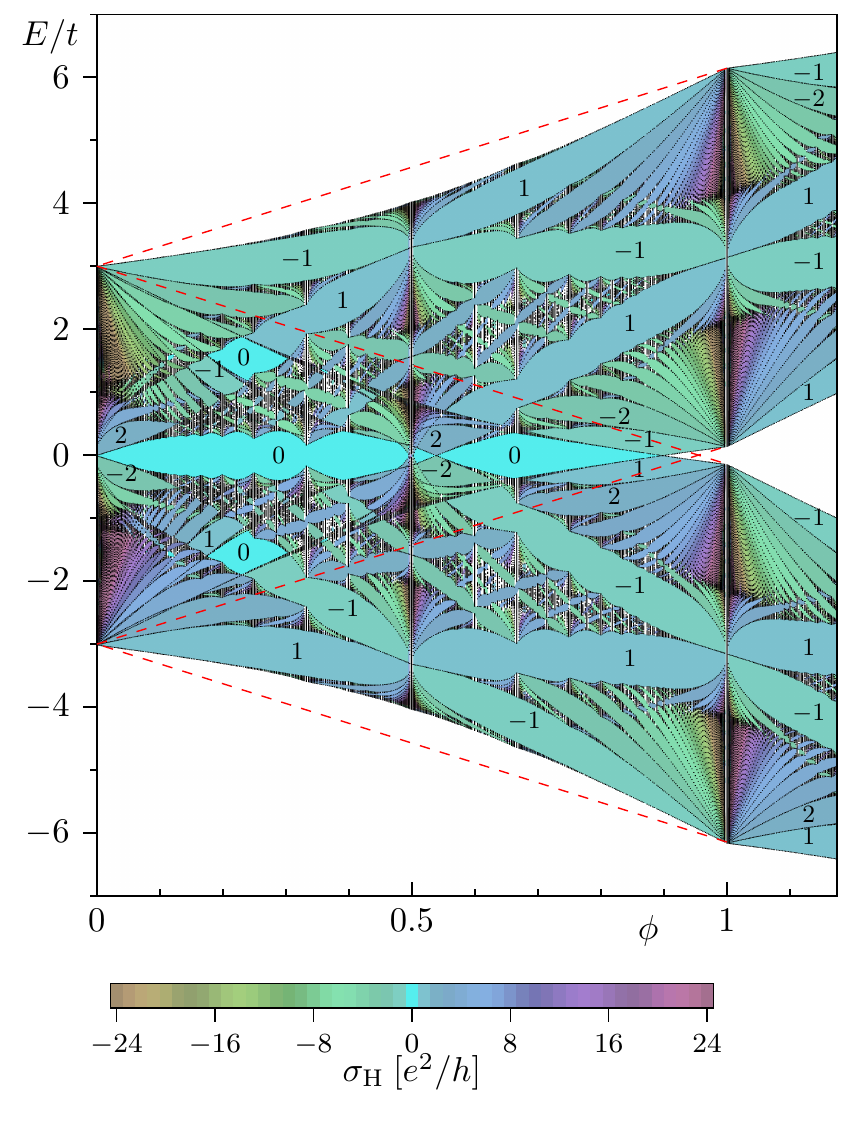}
\caption{\label{fig_spectrum_zeeman}\coloronline Hofstadter butterfly spectrum for the model described by Hamiltonian \eqref{eqn_hamiltonian} with Zeeman effect, $\lambdaZ/t=0.5$ and $\tI=\tR=0$. The bulk bands are in black. The colors and numbers indicate the quantized Hall conductivity in units of $e^2/h$. The diagonal dashed lines indicate the outline of the butterfly (i.e., $E/t=\pm 3\pm2\pi\phi\lambdaZ/t$) for the two spin components.}
\end{figure}

The Zeeman term models an on-site spin-splitting effect that is always present in real materials subjected to magnetic fields. Studying Hamiltonian~\eqref{eqn_hamiltonian} with $\lambdaZ\not=0$ and $\tR=\tI=0$, we observe that the spin degeneracy is lifted by merely shifting the two spin states up and down in energy by $2\pi\lambdaZ\phi$. The spectrum is otherwise left invariant. With respect to the Hofstadter butterfly represented in Fig.~\ref{fig_spectrum_spinless}, this means that the spin-up and spin-down copies are ``skewed'' in opposite directions, as shown in Fig.~\ref{fig_spectrum_zeeman}. Because the lowest and highest bands of the spin-degenerate model are situated at approximately $E/t=\pm3$, the two spin components are completely separated above a certain flux value, approximately equal to $\phi=3/(2\pi\lambdaZ/t)$. Furthermore, the Zeeman shift breaks the periodicity of the butterfly spectrum along the $\phi$ axis.

\subsubsection{The Zeeman-induced weak QSH phase}
\label{sect_zeeman_flux_weakqsh}

\begin{figure*}[t]
\includegraphics{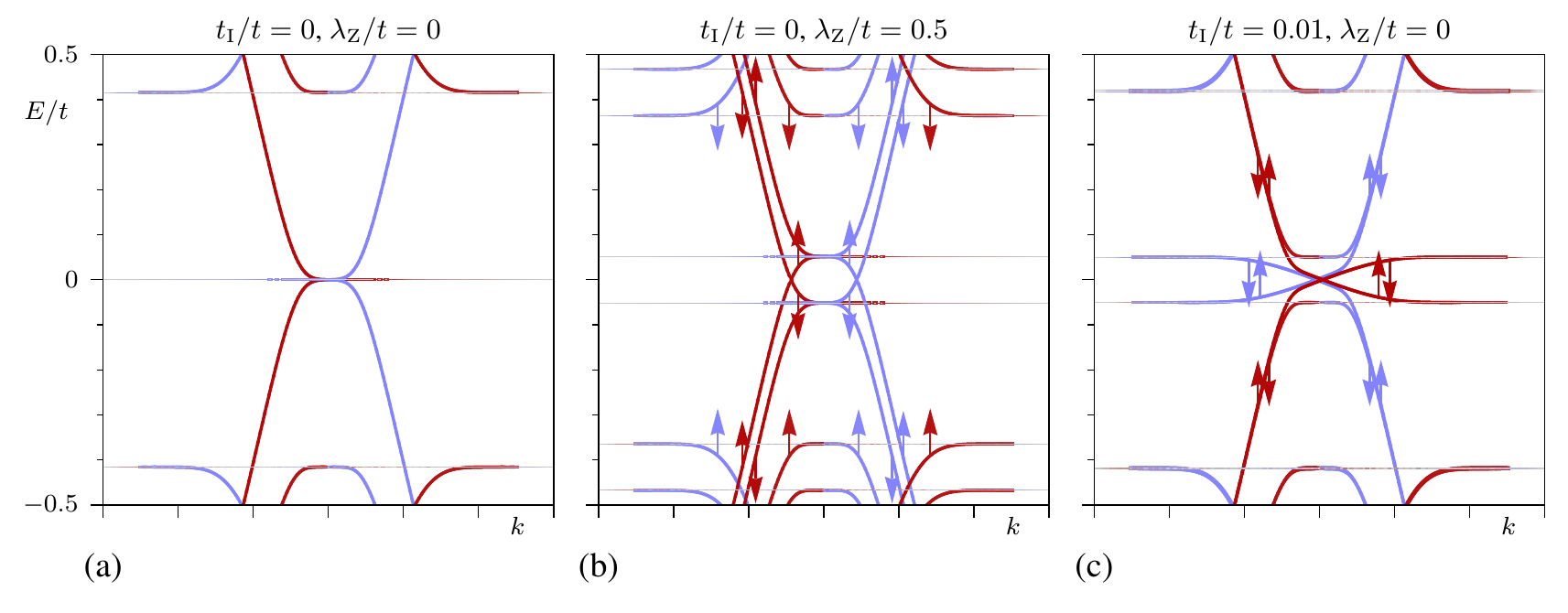}
\caption{\label{fig_dispersions_zeeman_or_iso}\coloronline  Comparison of dispersions at $\phi=1/61$ for (a) the spin-degenerate case, (b) the system with Zeeman effect ($\lambdaZ/t=0.5$) and (c) with ISO ($\tI/t=0.01$). The coupling constants $\lambdaZ$ and $\tI$ have been chosen such that the (weak-QSH) gap sizes at zero energy become equal.}
\end{figure*}

\par  As we already mentioned, the spectrum of the spin-degenerate model shows Landau levels for low values of the flux (cf. \sect\ref{sect_spindeg}). The Hall conductivities in the gaps are $\sigmaH/(e^2/h)=\ldots,-6,-2,2,6,\ldots$. If we concentrate on the bulk gaps around zero energy, we observe one and then three doubly degenerate  edge states (see Fig.~\ref{fig_dispersions_zeeman_or_iso}(a)). When the Zeeman splitting is nonzero, the two copies of these states shift up and down, as shown in Fig.~\ref{fig_dispersions_zeeman_or_iso}(b) for $\phi=1/61$. Assuming that the splitting is sufficiently large, a new gap opens at zero energy, where the original band was situated. The spin-up and spin-down edge states now connect to bulk bands that have been shifted up and down, respectively. As a consequence, in the newly opened gap at zero energy, we have $N_{\spinup}=-1$ and $N_{\spindown}=+1$. Thus, the charge Hall conductivity vanishes, and the spin Hall conductivity is equal to $\sigmaspH=-2e/4\pi$. Based on these values, we find that this state is a weak QSH phase.

\par Comparing this state to the one generated by ISO (see Fig.~\ref{fig_dispersions_zeeman_or_iso}(c)), we observe the similarity between them. In both cases, the Landau levels for the two spin components shift in opposite directions. The difference between the two terms lies in the different values of the shift: The Zeeman shift $\pm2\pi\phi\lambdaZ$ is linear in the flux, so that it vanishes for $\phi\to0$, while the Landau-level shift induced by the ISO coupling equals $\pm3\sqrt{3}\tI$.\cite{KaneMele2005PRL95-14} A recent study shows that a flux-independent Zeeman term (called an exchange term) may also generate the QSH state at zero magnetic field.\cite{YangEA2011PRL}

\par The similarity between the Zeeman and ISO terms may be understood at a formal level, in terms of the respective linearized (low-energy) Hamiltonians: The linearized Zeeman term acts as $\Psi^\dagger(\sigma_z^{\spinup\spindown}\otimes\idmat^{AB}\otimes\idmat^{KK'})\Psi$, where the factors indicate the proper spin, sublattice pseudospin, and valley pseudospin, respectively. On the other hand, the ISO coupling acts as $\Psi^\dagger(\sigma_z^{\spinup\spindown}\otimes\sigma_z^{AB}\otimes\sigma_z^{KK'})\Psi$. \emph{A priori}, these Hamiltonians act differently on the eight-component field vector $\Psi=(\psi_{\sigma,\tau,\xi})_{\sigma=\spinup\spindown,\tau=AB,\xi=KK'}$. In the lowest Landau level, the four components $\psi_{\spinup,A,K},\psi_{\spindown,A,K},\psi_{\spinup,B,K'},\psi_{\spindown,B,K'}$ vanish.\cite{Goerbig2011RMP} Substituting the remaining field components into the linearized Hamiltonians yields equality up to a sign,
\begin{align}
  &\Psi^\dagger\left(\sigma_z^{\spinup\spindown}\otimes\idmat^{AB}\otimes\idmat^{KK'}\right)\Psi\nonumber\\
  &{}=\psi^\dagger_{\spinup,A,K'}\psi_{\spinup,A,K'} - \psi^\dagger_{\spindown,A,K'}\psi_{\spindown,A,K'}\nonumber\\
  &\qquad{}+ \psi^\dagger_{\spinup,B,K}\psi_{\spinup,B,K} - \psi^\dagger_{\spindown,B,K}\psi_{\spindown,B,K}\nonumber\\
  &{}=-\Psi^\dagger\left(\sigma_z^{\spinup\spindown}\otimes\sigma_z^{AB}\otimes\sigma_z^{KK'}\right)\Psi,%
  \label{eqn_zeeman_iso_equivalence}%
\end{align}
which shows that the Zeeman effect and the ISO coupling act equivalently on the lowest Landau level. We remark that this reasoning is only valid  at  this specific Landau level: For higher Landau levels, all the eight fields are present, and the two terms become inequivalent.

\subsubsection{The spin-imbalanced quantum Hall phases}

\par So far, most of the studies on the QSH effect have generally concentrated on the behavior of the system at zero energy. However, very interesting features also emerge at nonzero energies  in the presence of external fields. For instance, in the low-flux regime, a Zeeman gap is formed at the $n=1$ Landau level (at $E/t\approx0.42$), as shown in Fig.~\ref{fig_dispersions_zeeman_or_iso}(b). Similarly to the zero-energy gap, this gap also shows a difference between the number of edge states with spin up and spin down components, $N_{\spinup}=1$ and $N_{\spindown}=3$, which may be understood from the values above and below the corresponding Landau level of the spinless model. Again, the spin Hall conductivity is nonzero, $\sigmaspH=-2\tfracs{e}{4\pi}$, but the difference with the weak quantum spin Hall gap is that the charge Hall conductivity is also nonzero, $\sigmaH=4\tfracs{e^2}{h}$, as  can be deduced from Fig.~\ref{fig_dispersions_zeeman_or_iso}(b). Furthermore, the edge states in this gap all propagate in the same direction (i.e., they are chiral), thus providing robustness against disorder. Here, we refer to this phase as the \emph{spin-imbalanced quantum Hall phase}.\cite{GoldmanEA2012}

\begin{figure}[t]
\includegraphics{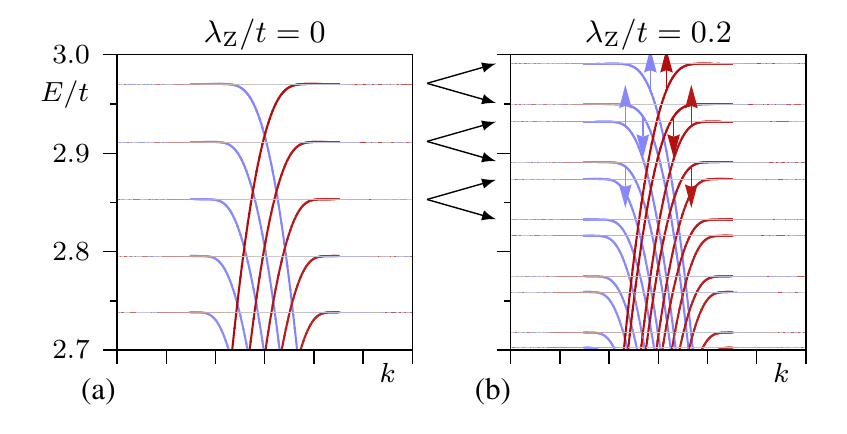}
\caption{\label{fig_zeeman_high_e}\coloronline  High-energy spectra for the model without spin-orbit couplings ($\tR=\tI=0$) at $\phi=1/61$. (a) spectrum for the spin-degenerate model ($\lambdaZ=0$), with quantum Hall gaps  corresponding to $\sigmaH=2,4,6,\ldots$  (in units of the conductivity quantum). (b) Spectrum with Zeeman effect ($\lambdaZ=0.2t$), showing the spin-filtered quantum Hall state with $N_\uparrow=1$ and $N_\downarrow=0$. The other gaps show (alternatingly) quantum Hall phases ($N_\uparrow=N_\downarrow$) and spin-imbalanced quantum Hall phases ($\abs{N_\uparrow}=\abs{N_\downarrow}+1$).}
\end{figure}

\par A special instance of the spin-imbalanced quantum Hall phase may occur if edge states of one of the spin-components are absent, i.e., either $N_{\spinup}=0$ or $N_{\spindown}=0$, in which case one speaks about the \emph{spin-filtered quantum Hall phase}. In this state, the magnitudes of the charge Hall and spin Hall currents are therefore equal when expressed in units of their respective conductance quanta. The spin-filtered quantum Hall phase appears at high energies and low flux in the presence of Zeeman coupling, as we now explain. In the absence of Zeeman coupling, we observe equally-spaced Landau levels, each of which has Chern number $-2$, see Fig.~\ref{fig_zeeman_high_e}(a). If the Zeeman term is present, the gap that forms between the two copies of the original highest-energy Landau level exhibits a spin-filtered quantum Hall phase, characterized by the presence of edge states of only one spin component; in this case $N_{\spinup}=1$ and $N_{\spindown}=0$, see Fig.~\ref{fig_zeeman_high_e}(b). The other gaps visible in Fig.~\ref{fig_zeeman_high_e}(b) are spin-imbalanced and ordinary spin-degenerate quantum Hall gaps, alternatingly. Let us mention that the spin-filtered and spin-imbalanced quantum Hall phases are ubiquitous in systems with Zeeman-split Landau levels, e.g., the quantum Hall plateaus corresponding to odd-integer filling factors in GaAs/AlGaAs heterostructures,\cite{DasSarmaPinczuk2008} or in HgTe quantum wells, which have strong Zeeman effect and are thus ideal candidates for observation of these phases.\cite{BeugelingEA2012PRB85}

\subsection{Rashba spin-orbit coupling}
\label{sect_rashba}%
\begin{figure}[t]
\includegraphics{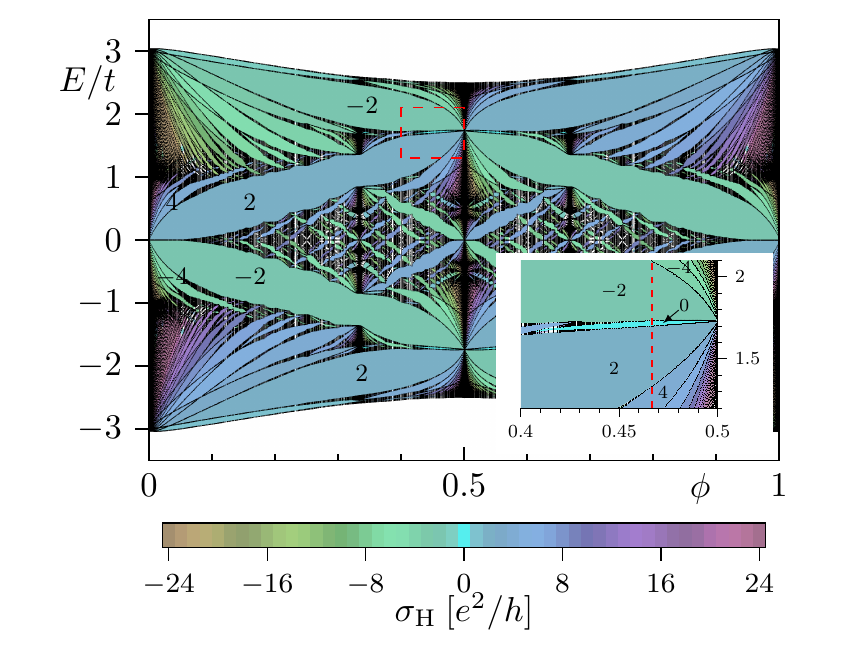}
\caption{\label{fig_spectrum_rashba}\coloronline Hofstadter butterfly spectrum (with RSO coupling, $\tR/t=0.2$). The inset shows a magnification of the region surrounded by the dashed lines. In the inset, the dashed line indicates the flux value $\phi=7/15$ discussed in the text. The colors and numbers indicate the quantized Hall conductivity in units of $e^2/h$.}
\end{figure}

\begin{figure*}[t]
\includegraphics{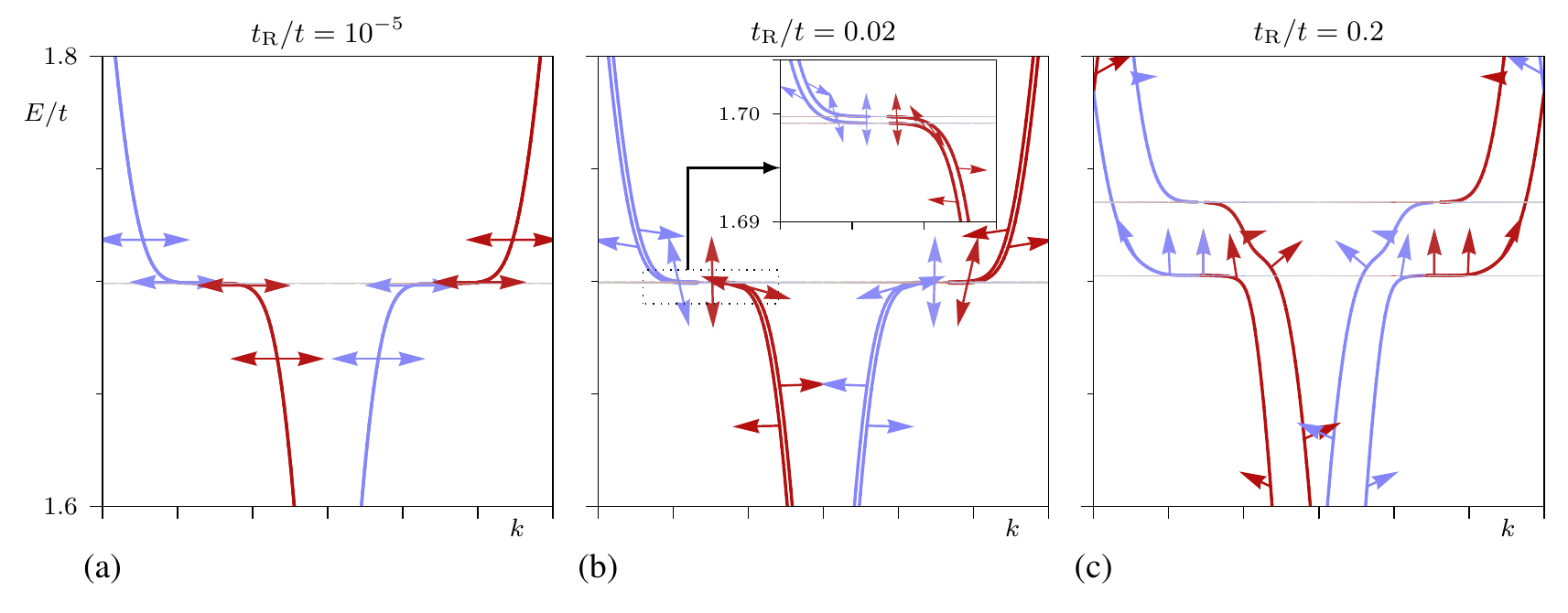}
\caption{\label{fig_dispersions_rashba}\coloronline  Dispersions at $\phi=7/15$,  showing the rotation of spin directions as the Rashba coupling $\tR$ is increased.}
\end{figure*}

\par The RSO coupling differs in an essential way from the ISO coupling and the Zeeman effect, in the sense that the hopping involves a spin flip. In other words, the spin matrices involved in the Rashba hopping are off-diagonal (i.e., $\sigma_x$ and $\sigma_y$), whereas the other terms in the Hamiltonian involve diagonal spin matrices. This has a profound effect on the spin structure of the system:  If the RSO interaction is nonzero, mixing occurs between the spin components and we find that the spin orientation of the eigenstates is generally site-dependent. The spin direction is confined to the two-dimensional plane and is perpendicular to the momentum: Aside from the out-of-plane component ($z$-direction), the component perpendicular to the cylinder edge may also be nonzero, while the component parallel to the edge  always vanishes. In other words, the helicity $(\sigma_x,\sigma_y,\sigma_z)\cdot\hbar\vec{k}$ of the (edge) states remains zero. This effect is caused by the nature of the Rashba coupling in the Hamiltonian \eqref{eqn_hamiltonian}, which involves the outer product of the spin vector $\vecsigma$ with the bond vectors $\vec{d}_{jk}$, so that the spin is always perpendicular to the hopping direction. Because the edge-state spins are no longer in the up or down state, the Chern numbers for the respective spin components are no longer well-defined. Thus, the spin conductivity no longer takes quantized values. Nevertheless, the spin Chern number remains a well-defined integer-valued topological invariant,\cite{ShengEA2006} which can be used as a tool to distinguish between trivial and nontrivial gaps.

\par The second difference with the ISO coupling is that the zero-flux spectrum remains gapless and the spin degeneracy is lifted by the RSO coupling (considering $\tR\not=0$ and $\tI=\lambdaZ=0$). Indeed, upon setting $\tR\not=0$, the two Dirac cones from the spinless model (at the special points $K$ and $K'$) are broken into four cones each: A central isotropic  cone at $K$ or $K'$, and three anisotropic ``satellite'' cones around it. This effect is known as trigonal warping.\cite{ZareaSandler2009,VanGelderenMoraisSmith2010} Between the main cones and the satellite ones, there is a Van Hove singularity at low energy, which scales as $\pm(\tR/t)^3$ for small $\tR/t$. The Berry phases associated to the main and satellite cones are $-\pi$ and $\pi$, respectively, which add up to the Berry phase $2\pi$ of the Dirac cones in the spin-degenerate model ($\pi$ for each spin component). \cite{[{For bilayer graphene, which is characterized by a Hamiltonian that is formally equivalent to $H_\mathrm{NN}+H_\mathrm{R}$, these Berry phases have been derived in }][{}]DeGailEA2012}
For energies $\abs{E/t}<(\tR/t)^3$, the low-flux Landau level spectrum is characterised by two different sets of Landau levels with Chern numbers $2$ and $6$. Outside this regime, there are only twofold-degenerate Landau levels, because the satellite cones can no longer be resolved. Around $\abs{E/t}\approx(\tR/t)^3$ there is a crossover regime where the sixfold-degenerate bands split into three, each one being twofold degenerate.

\par Adding a nonzero Rashba term to the spin-degenerate model discussed in \sect\ref{sect_harper_hofstadter} will cause a spin-splitting of some bulk bands in the Hofstadter butterfly, see Fig.~\ref{fig_spectrum_rashba}. In particular, at low energies, the Rashba term splits the fourfold-degenerate Landau levels into twofold-degenerate Landau levels. In the limit $\phi \rightarrow 0$, we also note that the spectrum indeed remains gapless and particle-hole symmetric. However, this is not true for other Dirac regimes in the spectrum. For instance, at $\phi=7/15$ and $E/t\approx1.7$ (see the inset of Fig.~\ref{fig_spectrum_rashba}), the RSO coupling splits the central Landau level, and the resulting gap exhibits two counterpropagating edge modes on each edge. However, the spin Hall conductivity is undefined because the spin direction varies as a function of the Fermi energy. Nevertheless, at each energy inside the gap the spins of the two edge states are approximately equal, so that the spin current nearly vanishes.  Besides, the charge Hall conductivity is exactly zero when the Fermi energy is located in this gap.

\par The RSO coupling has drastic effects on the edge states, as their spins no longer align perpendicularly to the sample plane, but get an additional in-plane component perpendicular to the edge. Besides, the direction of the edge-state spins depends on the Fermi energy. This observation distinguishes the Rashba effect from the Zeeman effect in a tilted field, which puts all edge-state spins in the same direction. In addition, the in-plane component is opposite for edge states at the opposite edges, whereas the perpendicular component is the same. Importantly, due to the dependence of the spin direction on the Fermi energy, the latter can be conveniently used for spin manipulations (in addition to the amplitude of the Rashba coupling itself).

\par We illustrate this phenomenon around the Rashba-split bulk band at $\phi=7/15$ and $E/t\approx1.7$. In Fig.~\ref{fig_dispersions_rashba}(a), we show that in the limit of $\tR\to 0$, the eigenstates are in-plane (indicated as ``left'' and ``right'' in the figure, meaning that the spin direction is $\pm\hat{y}$). This result shows that for infinitesimal RSO coupling, it is more natural to decompose the states in terms of the eigenstates of the Pauli matrix $\sigma_y$ than those of $\sigma_z$. If the Rashba coupling is then increased, the gap opens, and in the vicinity of the bulk bands, the edge states tend to rotate to the vertical direction, see Fig.~\ref{fig_dispersions_rashba}(b). In addition, we observe that not all spins have the same length. This phenomenon occurs because the spins displayed in this plot represent the expectation values of the spin components (see the Appendix). Since the spin direction depends on the lattice position ($y$ coordinate), the length of this expectation value may be less than unity.

\par We expect that the canting of the spins is most easily observed in gaps with a single edge state on each edge. For instance, in Ref.~\onlinecite{GoldmanEA2012}, it has been demonstrated that the RSO coupling cants the spin in a spin-filtered gap generated by the Zeeman effect. In this setting, the spin textures may be controlled at will by tuning the coupling parameters and the Fermi energy.

\begin{figure}[!b]
\center\includegraphics[scale=1]{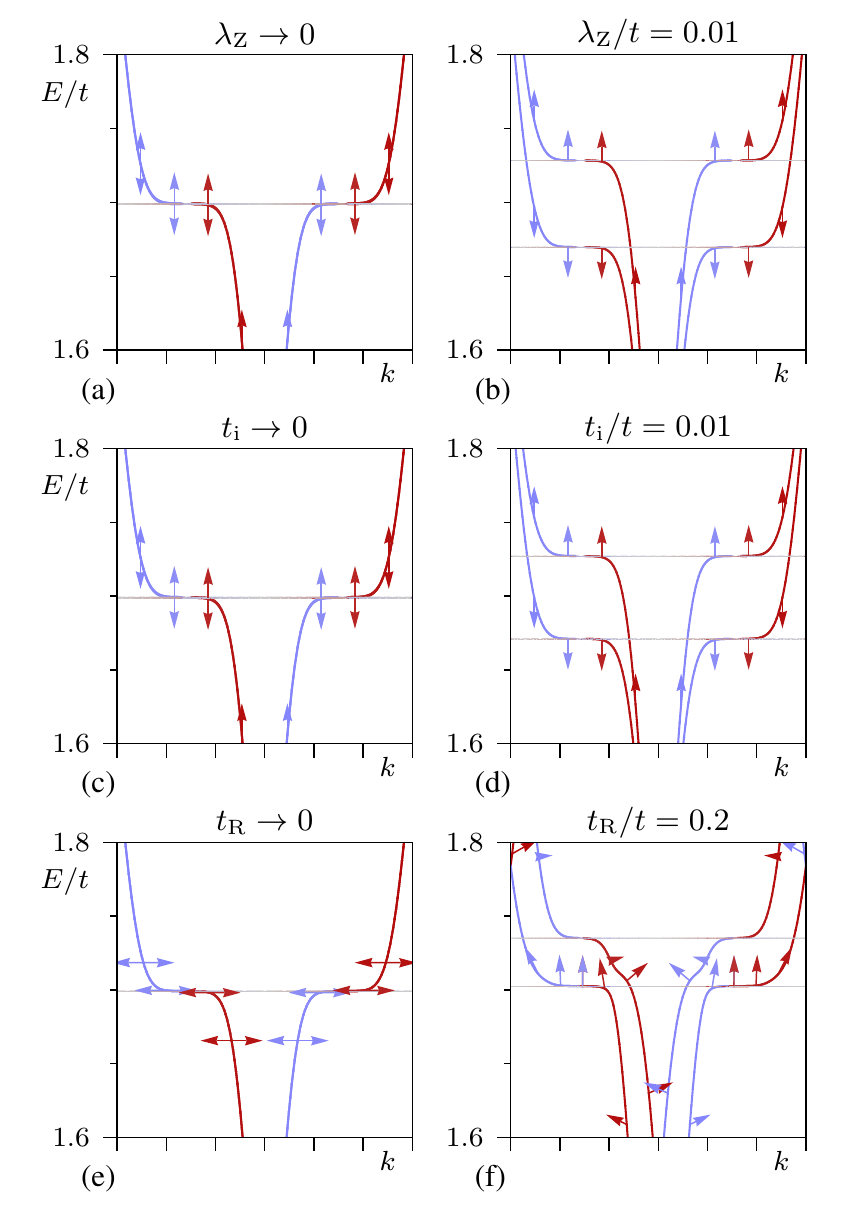}
\caption{\label{fig_splitting_examples}\coloronline Spin splitting in an example at flux $\phi=7/15$ and near $E/t=1.7t$. Zeeman effect: (a) $\lambdaZ\to0$ and (b) $\lambdaZ=0.01t$; ISO coupling: (c) $\tI\to0$ and (d) $\tI=0.01t$; RSO coupling: (e) $\tR\to0$ and (f) $\tR=0.2t$. }
\end{figure}

\subsection{Comparison between the low-flux limit $\phi \approx 0$ and the ``half-flux" regime $\phi\approx 1/2$}
\label{sect_halfflux}%

\par As shown before in the low-flux limit $\phi\to 0$, the ISO coupling and the Zeeman effect affect the zero-energy modes  equivalently in the sense that they both open a weak QSH gap. Indeed, it has been shown that for this mode, the ISO coupling and Zeeman effect are formally equivalent.\cite{Goerbig2011RMP} On the other hand, the RSO coupling does not open a gap at zero energy. This fact hampers the comparison of the RSO coupling to the aforementioned two terms in this energy regime.

\par From the Hofstadter butterfly shown in Fig.~\ref{fig_spectrum_spinless}, one observes that the Landau-level structure in the low-flux, low-energy limit is not unique. In fact, similar structures appear wherever the shape of the dispersion is characterized by Dirac cones. For example, at $\phi=1/2$ and for energies close to $E/t=\pm \sqrt{3}$, the dispersion can be approximated by a Dirac dispersion of the form $\abs{E/t}=\sqrt{3}\pm\sqrt{\tfrac{3}{2}}\abs{\vk}-\tfrac{1}{4}\sqrt{3}\abs{\vk}^2+\mathcal{O}(\abs{\vk}^3)$, where $\vk$ is the momentum relative to the position of one of the two Dirac cones. The linear term is responsible for the square-root behavior of the (fourfold-degenerate) Landau levels for flux values close to $\phi=1/2$. The quadratic term causes the asymmetry of the Landau level spectrum with respect to the energy $\abs{E/t}=\sqrt{3}$. Furthermore, we observe that the energy $\abs{E_{\mathrm{LL},0}/t}\approx\sqrt{3}-(\pi/3)\abs{\phi-1/2}$ of the central Landau level is linear in the flux rather than constant.

\par The approximate linear dependence of the spectrum  $E(\vk)$ at $\phi=1/2$ and  $\abs{E/t}\approx \sqrt{3}$  suggests that the Zeeman and ISO terms should behave similarly in this regime, in direct analogy to the case $\phi, E \approx 0$ discussed in \sect\ref{sect_zeeman_flux_weakqsh}. Although the energy of the lowest Landau level is no longer constant in $\phi$, the approximation holds, thus leading to a similar behavior with regard to the opening of the gaps, as shown in Figs.~\ref{fig_splitting_examples}(a--d) for the flux value $\phi=7/15$ which lies close to $\phi=1/2$. The Zeeman effect and ISO coupling split the central Landau level at $E/t\approx1.7$ and open up a weak QSH gap, similar to the zero mode in the low-flux regime. As discussed in \sect\ref{sect_rashba}, the RSO coupling splits the central Landau level, unlike the case for $\phi\to0$, see Figs.~\ref{fig_splitting_examples}(e,f). Although there are edge states inside the gap, it is a trivial state, because the spin Hall conductivity approximately vanishes. This comparison shows an important difference between Zeeman coupling and ISO coupling on one hand, and the Rashba coupling on the other hand: The former two create the weak QSH phase, while the latter does not. In the following section, we show that in the presence of Zeeman or ISO coupling, the Rashba coupling tends to destroy the weak QSH phases generated by the former effects.


\section{Competition and phase transitions}
\label{sect_competition_and_phasetrans}%
\par After analyzing the effects of the three terms $H_\mathrm{I}$, $H_\mathrm{R}$, and $H_\mathrm{Z}$ in Hamiltonian~\eqref{eqn_hamiltonian} independently, a natural question arises: what effects emerge when these terms are combined? Here, we study the competition featuring two terms, by tuning the ratio between their corresponding coupling constants. Tuning these coupling constants generally leads to opening and closing of gaps, so that the topological nature of the gaps may change. In that case, we deal with \emph{topological phase transitions}. We recall that Kane and Mele\cite{KaneMele2005PRL95-14} have studied the phase transition from a (time-reversal symmetric) QSH state to a trivial state, by increasing the ratio $\tR/\tI$. Here, we extend this study and we explore the more exotic phase transitions that are realized in the presence of a magnetic field and the Zeeman effect.

\subsection{Phase transitions driven by the ISO coupling}
\label{sect_iso_phase_transition}%
\begin{figure*}[!t]
\includegraphics{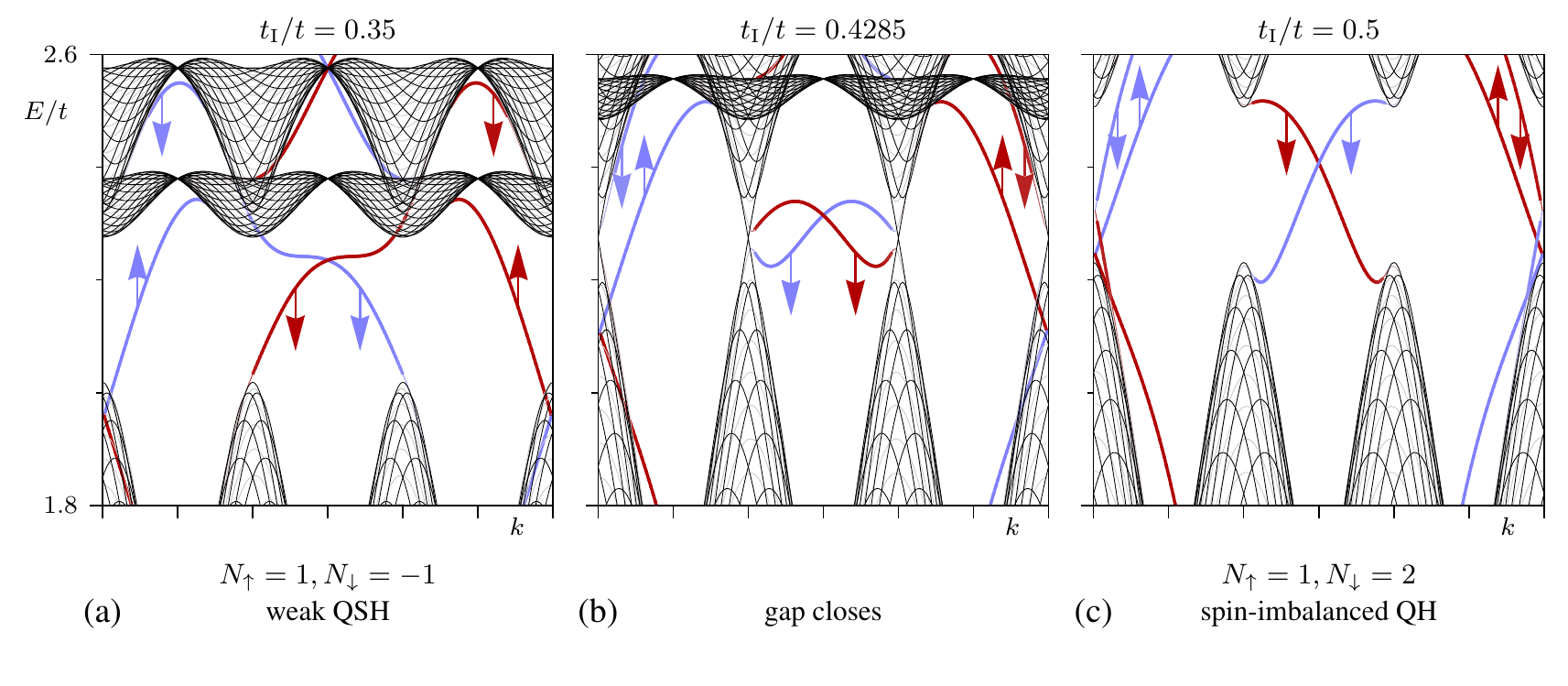}
\caption{\label{fig_phase_transition_iso}\coloronline Spectra for $\phi=1/3$, with (a) $\tI/t=0.35$, (b) $\tI/t=0.4285$, (c) $\tI/t=0.5$. In (a), we observe a (weak) QSH gap, where $N_{\spinup}=-N_{\spindown}=1$. (b) By increasing the ISO coupling constant, the bulk gap closes for $\tI/t\approx0.4285$. (c) If $\tI$ is increased further, the gap reopens, but with a different number of edge states, namely $N_{\spinup}=1$ and $N_{\spindown}=2$, so that it enters a spin-imbalanced phase. Thus, the system undergoes a topological phase transition from a helical phase (weak QSH) to a chiral phase (spin-imbalanced QH).}
\end{figure*}

\par Let us analyze the interplay between the ISO coupling and the magnetic field by investigating a gap at a fixed flux value, in absence of the Zeeman effect and of the Rashba coupling. By tuning the value of the ISO coupling amplitude $\tI$, we can close the gaps and obtain different topological phases at both sides of the transition. In Fig.~\ref{fig_phase_transition_iso}, we have shown the dispersions for $\phi=1/3$, for values of $\tI$ below, at, and above the values where the gap around $E/t=2.3$ closes. Around this energy, there are three bulk bands: Two (overlapping) ones above and one below the gap. For values of $\tI<t_{\mathrm{I},0}\approx0.4285t$, below the critical value, we observe the weak QSH phase, i.e., two counterpropagating edge modes with opposite spins on each edge, $N_{\spinup}=1$ and $N_{\spindown}=-1$. The spin-down pair of edge modes connects the top and bottom bulk bands visible in this plot, and the spin-up pair connects the middle band to a lower band. For $\tI=t_{\mathrm{I},0}$, the gap disappears: The bulk bands touch each other in three points at $E/t\approx2.3$, and two of these points are connected by the spin-down edge states in the middle of the figure. If the value of $\tI$ is increased further, a gap opens again, and a new pair of edge states (with spin down) appears. Moreover, the existing spin-down edge states have inverted their direction of propagation. Thus, for $\tI>t_{\mathrm{I},0}$, we have a spin-imbalanced quantum Hall state with $N_{\spinup}=1$ and $N_{\spindown}=2$. Note that only the spin-down states are modified in this phase transition, while the spin-up states remain the same.

\par In Ref.~\onlinecite{GoldmanEA2012}, this phase transition was investigated in the presence of a nonzero Zeeman coupling in addition to the ISO coupling. Here, we emphasize that the Zeeman effect is actually superfluous and that this transition is driven exclusively by the ISO coupling. Variation of $\lambdaZ$ has no qualitative effect on the transition, but only shifts the energies of the bands and gaps. Within a large range of values for $\lambdaZ$, we obtain an identical phase transition as in Fig.~\ref{fig_phase_transition_iso}. The critical value $t_{\mathrm{I},0}$ of the ISO strength is independent of $\lambdaZ$.

\par The phase transition presented here differs in a fundamental way from a crossing of Landau levels. In the latter case, the Chern numbers associated to the Landau levels do not change at the transition, and as a consequence the number of edge states inside the gap between them remains unmodified as well. For the phase transition presented here, the number of edge states and consequently the Chern numbers do change at the transition. Focusing on the spin-down edge states, we observe that $N_{\spindown}=-1$ for $\tI<t_{\mathrm{I},0}$ and $N_{\spindown}=2$ for $\tI>t_{\mathrm{I},0}$, a difference of $\Delta N_{\downarrow}=3$. The Chern numbers of the bands below and above the gap are $1$ and $1$ in the former case and $-2$ and $4$ in the latter. The sum of these Chern numbers is unchanged, as required by the bulk-boundary correspondence, \eqn\eqref{eqn_bbc}.

\par The question arises as to whether the difference in the number of edge states at both sides of the transition may be predicted. \emph{A priori}, this difference cannot be predicted since it is the result of a complicated interplay of the ISO coupling and the magnetic flux. However, the difference is always a multiple of the denominator $q$ of the flux $\phi=p/q$. This result may be understood from the fact that at a given flux value $\phi=p/q$, the Chern numbers of all the Hofstadter bands obey $C_n\equiv c\pmod q$. Here, $c$ is the \emph{modular multiplicative inverse} of $p$ modulo $q$, defined as the unique integer $c$ ($0\leq c<q$) such that $cp\equiv1\pmod q$.\footnote{This result has been verified empirically for the spinless model. The observation that $C_n\equiv c\pmod q$ for all $n$ implies automatically that $cp\equiv1\pmod q$, assuming that the St\v{r}eda formula and the bulk-boundary correspondence hold. These results are valid more generally for the model described by Hamiltonian~\eqref{eqn_hamiltonian} as long as the two spin components are decoupled.} Consequently, the difference between two possible values of the Chern numbers is always a multiple of $q$, which proves the aforementioned claim. This property may also be understood from the superlattice structure: The magnetic field produces $q$ copies of the original unit cell, and each of them reacts in the same way to the perturbation leading to the phase transition.

\begin{figure}[t]
\includegraphics{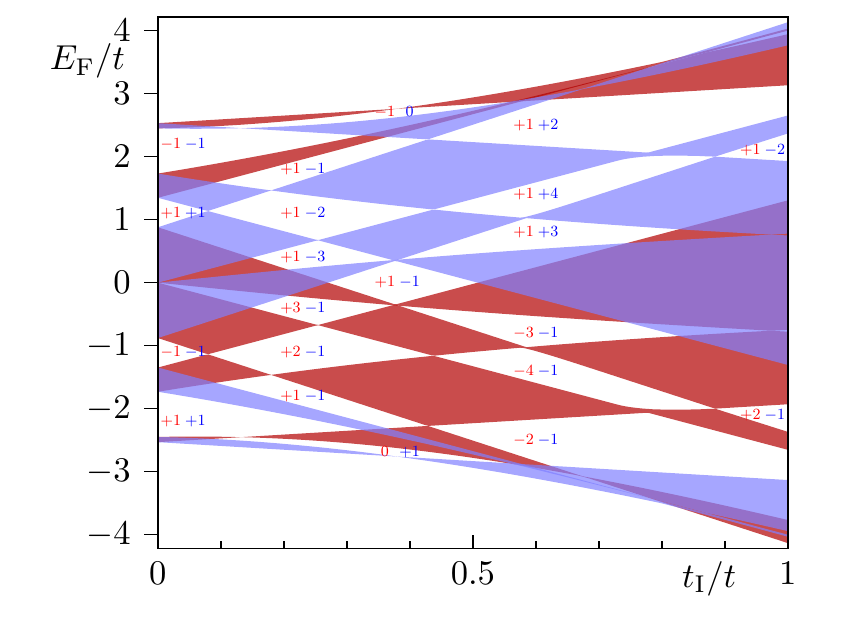}
\caption{\label{fig_phase_diagram_vs_iso}\coloronline  Phase diagram as a function of the ISO coupling $t_\mathrm{I}$ and the Fermi energy $E_\mathrm{F}$ for the fixed flux value $\phi=1/3$. In the shaded regions, the system is metallic. The colors red (dark gray) and blue (light gray) distinguish the spin-up and -down components, respectively. In the white regions, the bulk is insulating, and the edges conduct. There, the pairs of numbers indicate the number of edge states $N_{\spinup}$ and $N_{\spindown}$ for spin up and down, respectively.}
\end{figure}

\par In Fig.~\ref{fig_phase_diagram_vs_iso}, we show the phase diagram as a function of the ISO coupling $t_\mathrm{I}$ and the Fermi energy $E_\mathrm{F}$ with the flux value $\phi=1/3$ held fixed. From this figure, we can easily read off the values of $\tI$ and the energies $E_\mathrm{F}$ where the gaps close and the topological phase transitions occur. A few points must be noticed: First of all, we clearly identify the weak-QSH phase at half-filling for $\tI/t=0.4$. The (spin-degenerate) QH phases are progressively destroyed for $\tI/t<0.3$. For $\tI/t>0.3$, we get spin-filtered QH [e.g., $(N_{\spinup},N_{\spindown})=(-1,0)$], spin-imbalanced QH [e.g., $(N_{\spinup},N_{\spindown})=(+1,+2)$] and the weak QSH phase. In addition, there are also more exotic phases where the edge states are neither chiral nor helical, e.g., $(N_{\spinup},N_{\spindown})=(+1,-2)$. The spin-degenerate phases do not arise for $\tI/t>0.3$. Secondly, the gaps are large (of the order of $t$) in wide regions of the parameters, which is favorable for the detection of the transitions in cold-atoms experiments. In addition, the property that the number of edge states changes by a multiple of $3$ at each phase transition is clearly observed in Fig.~\ref{fig_phase_diagram_vs_iso}. Finally, the energies where the phase transitions occur may be shifted by tuning the strength of the Zeeman effect. Since the spin-up and spin-down components are uncoupled, the Zeeman effect will shift all red (blue) areas in Fig.~\ref{fig_phase_diagram_vs_iso} up (down) by a fixed amount of energy. We note that the Zeeman effect alone does not induce the same type of phase transitions as the ISO coupling. The Zeeman effect can only close and open gaps between bands of different spin components, and does therefore not modify the Chern numbers of the bands, whereas the ISO coupling can also close gaps between two bands of the same spin component. Nevertheless, tuning the Zeeman coupling strength $\lambdaZ$ allows one to modify the nature of the phase transitions driven by ISO coupling.

\subsection{Intrinsic spin-orbit coupling and Zeeman effect for $\phi\not=0$}
\label{sect_iso_zeeman_flux}%

\begin{figure}[t]
\includegraphics{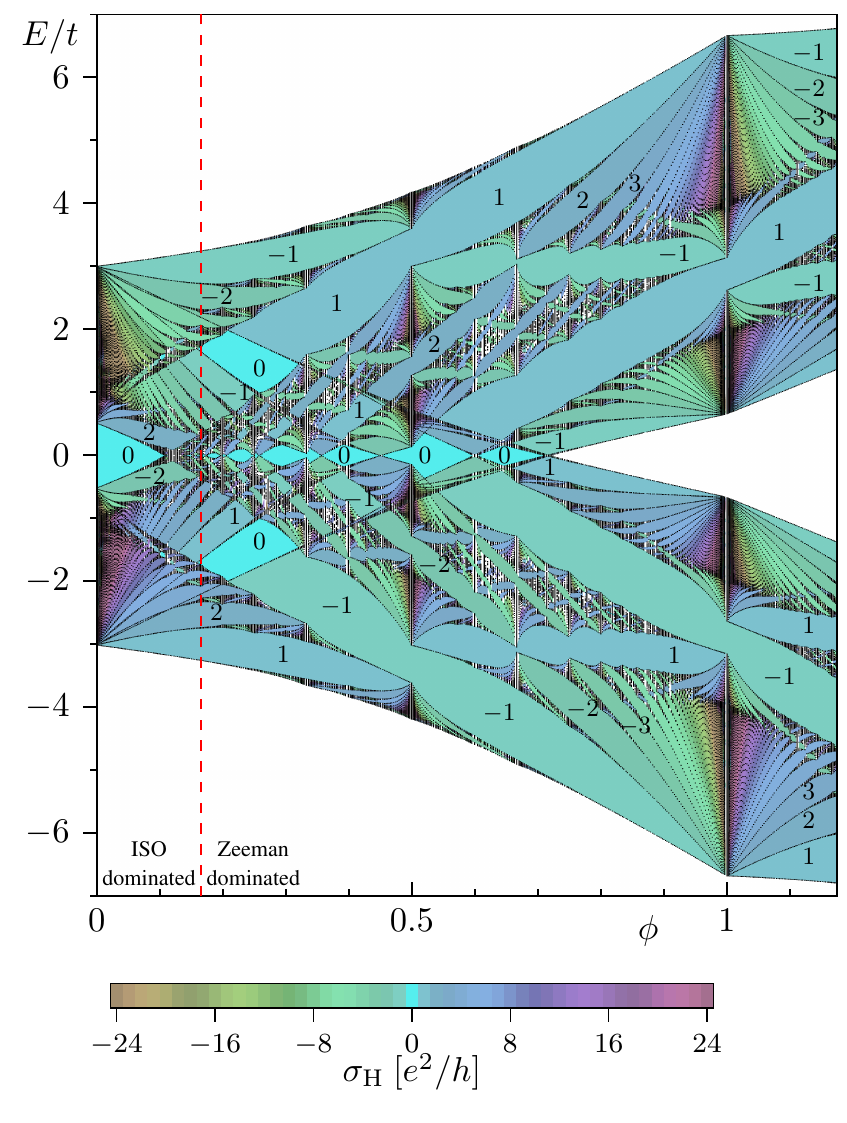}
\caption{\label{fig_spectrum_zeeman_iso}\coloronline Hofstadter butterfly spectrum with ISO coupling ($\tI/t=0.1$) and Zeeman effect ($\lambdaZ/t=0.5$). the colors and numbers inside the bulk gaps indicate the quantized Hall conductivity in units of $e^2/h$. The vertical dashed line indicates the flux value where the two effects are comparable in strength. At this value, the crossover from the ISO-dominated (low $\phi$) to the Zeeman-dominated (high $\phi$) regime takes place.}
\end{figure}

\par  If the Zeeman effect and ISO coupling are present simultaneously, the interplay between the two terms is governed by the value of the flux $\phi$, because the amplitude of the former is linear in $\phi$ while that of the latter is constant. Thus, in the butterfly spectrum illustrated in Fig.~\ref{fig_spectrum_zeeman_iso}, we observe the ISO-dominated regime at low flux (cf.\ Fig.~\ref{fig_spectrum_iso}), and the Zeeman-dominated regime at high flux (cf.\ Fig.~\ref{fig_spectrum_zeeman}). The crossover between these two regimes takes place where the two terms are comparable in strength, approximately where $4\pi\phi\lambdaZ\sim6\sqrt{3}\tI$, indicated by the vertical dashed line in Fig.~\ref{fig_spectrum_zeeman_iso} at $\phi\approx 0.165$. We remark that this crossover effect is present more generally in the situation where the Zeeman term competes with any other term with a constant amplitude: Due to its linear $\phi$-dependence, the Zeeman term will always dominate any other one at sufficiently high values of $\phi$. Naturally, this phenomenon only occurs in systems where the Zeeman splitting depends linearly on the magnetic flux $\phi$. In the presence of magnetic impurities, the Zeeman term may exhibit a nonlinear dependence on the applied magnetic flux, as discussed in Ref.~\onlinecite{BeugelingEA2012PRB85} for Mn-doped HgTe quantum wells, but this subject is beyond the scope of this paper.

\par Although the Zeeman effect dominates over the ISO coupling for $\phi\gg0.165$, the influence of the ISO coupling is still visible in the high-flux spectrum: At $\phi=1/2$ and $\phi=1$, there are gaps at $E/t\approx 3$ (see Fig.~\ref{fig_spectrum_zeeman_iso}), which are absent in the butterfly without the ISO coupling (see Fig.~\ref{fig_spectrum_zeeman}).  We note that the dispersions around these points are of Dirac type for $\tI=0$,  consequently, the ISO coupling opens these gaps in a way analogous to the gaps at zero flux and zero energy. However, the gaps at $\phi=1/2$ and $\phi=1$ are spin-filtered QH phases and not QSH, due to the subtle competition between the ISO coupling and the TRS-breaking effects.

\subsection{Rashba spin-orbit coupling and exchange term at $\phi=0$}
\label{sect_rashba_exchange}%

In two recent works,\cite{QiaoEA2010,*QiaoEA2012} it has been shown that the combination of RSO and an exchange field, similar in structure to the Zeeman term, leads to chiral edge states at zero energy. In addition, the problem has been investigated for a sodium/lithium iridate model which involves NN and NNN hopping terms.\cite{MurthyEA2012} In the latter and more complex setup, nontrivial QH gaps appear also away from half filling. Here, we consider the combined effects of the Zeeman and Rashba couplings, away from half-filling and in the absence of ISO coupling. We demonstrate that \emph{several} QH gaps appear in the absence of NNN hopping terms, without including any gauge field (i.e., the Peierls phases): Namely, non-trivial QH phases are produced in our model by setting $\phi=\tI=0$, $\tR \ne 0$ and by applying a flux-independent Zeeman, or exchange,\cite{QiaoEA2010,*QiaoEA2012,YangEA2011PRL} term
\begin{equation}
H_{\mathrm{ex}}= \gex \sum_jc_j^\dagger \sigma_z c_j,
\end{equation}
which differs from $H_{z}$ by the fact that the strength $\gex$ no longer depends on the flux $\phi$ (which we now set to zero). We stress that it is the simple association of a constant exchange term with a NN Rashba hopping term that leads to non-trivial topological phases, at half-filling but also at $E_{\mathrm{F}}\approx \gex$, as discussed in detail below.

\begin{figure}[t]
	\centering
	\includegraphics[width=1\columnwidth]{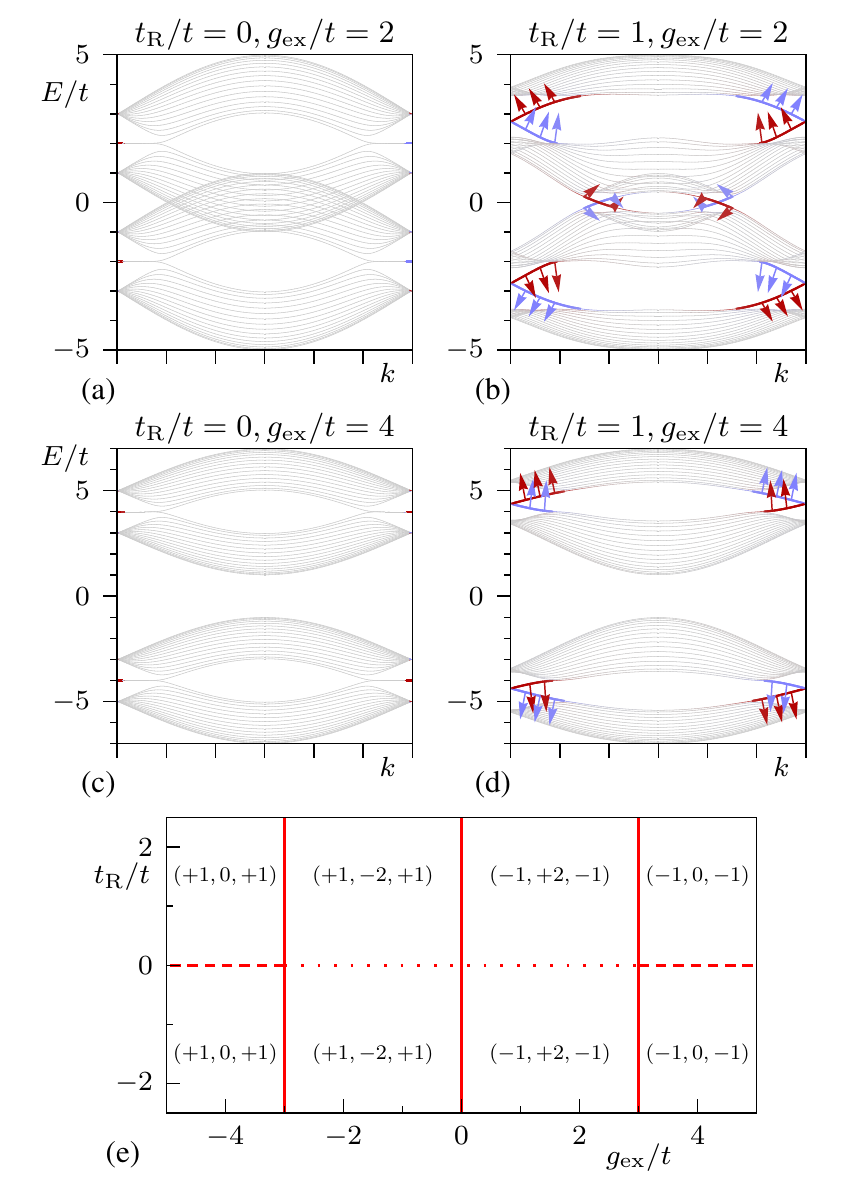}
	\caption{\label{fig-rashba-exch}\coloronline  (a-d) Energy spectra $E=E(k)$ for $\phi=\tI=0$. The Zeeman (exchange) and Rashba coupling strength are indicated. (e) Phase diagram in the parameter space defined by the coupling strengths $g_\mathrm{ex}$ and $t_\mathrm{R}$. The number triplets indicate the charge Hall conductivity in the lower, middle, and upper gaps, respectively. The solid lines indicate the phase transitions for the gap at zero energy. At the dashed lines, the lower and upper gaps close and the trivial gap at zero energy remains open [as in (c)], and at the dotted lines, all three gaps close [as in (a)]. At the dashed and dotted lines, the charge Hall conductivities in the gaps do not change.}
	\end{figure}

When the exchange term is added to the tight-binding model with strength $0<\gex < 3t$, the energy bands associated to the spin-up and spin-down components are shifted in opposite directions but overlap at half-filling. In this situation, which is illustrated in Fig.~\ref{fig-rashba-exch}(a), the system is a semi-metal for $E_{\text{F}}= \pm \gex$ and a metal otherwise. When the Rashba coupling is turned on, bulk gaps open not only at half-filling $E=0$ but also at $E= \pm \gex$ (cf.\ Fig.~\ref{fig-rashba-exch}(b)). We have computed the edge-state structures as well as the topological indices associated to the gaps: we obtain that all these gaps are related to non-trivial QH phases. More precisely, we find that the gaps at $E= \pm \gex$ host a single edge state per edge, the spin orientation of which is given by $\langle \sigma_z \rangle \approx \pm 1$: these gaps correspond to spin-filtered QH phases, with equal (resp.\ opposite) charge (resp.\ spin) Hall conductivities. We observe that the spin orientation of these isolated edge states is nearly vertical, which is due to the fact that the exchange term is important in this high-energy regime. At half-filling, the bulk gap hosts two edge states per edge, with same velocity and quasi-horizontal spin orientation. This bulk gap corresponds to a topological QH phase with $\sigmaH=2$ and a non-trivial spin orientation. Therefore the spin orientation of the edge states highly depends on the Fermi energy. We note that the gap openings occur for large exchange coupling, but for arbitrarily low Rashba coupling.

When the exchange strength $\gex > 3t$, the energy bands associated to the spin-up and spin-down components are well separated: a trivial gap opens at half-filling, while the Dirac points are shifted to the energies $E=\pm \gex$, cf.\ Fig.~\ref{fig-rashba-exch}(c). In this situation, the Rashba coupling still opens two non-trivial QH phases at $E_{\text{F}}=\pm \gex$, cf.\ Fig.~\ref{fig-rashba-exch}(d). Again, these phases are characterized by spin-filtered edge states with nearly vertical spin orientation. Therefore, for sufficiently large exchange coupling, it is necessary to set the Fermi energy away from half-filling in order to detect nontrivial topological phases. This behavior is summarized in the phase diagram of Fig.~\ref{fig-rashba-exch}(e): For nonzero Rashba coupling $\tR$, the Hall conductivity at zero energy is $\sigmaH=\pm2$ for $\abs{\gex/t}<3$ and zero otherwise. Thus, topological phase transitions of the zero-energy gap take place at the gap closings at $\gex/t=\pm3$. The lower and upper gaps (at $E_\mathrm{F}\approx \pm\gex$) always exhibit a Hall conductivity of $\pm1$, where the sign depends on that of $\gex/t$ but not on that of $\tR/t$. Inverting the sign of $\gex$ inverts the directions of the edge currents, as well as the $z$ component of the spins. Inverting the sign of $\tR$ inverts the in-plane component of the spins. Continuous variation of $\tR$ through zero changes the spin-direction continuously, where the spins tend to the $z$ direction in the limit $\tR/t\to0$. Thus, the gap closing at $\tR=0$ does not constitute a topological phase transition.

Let us emphasize the fact that the chiral QH phases presented here are not the consequence of the external magnetic flux (i.e., Peierls phases), since they are produced when $\phi=0$, and solely in the presence of the exchange and Rashba terms.  When $\phi \not= 0$, new QH gaps open in the system, altering the exchange-Rashba-induced phases presented in Figs.~\ref{fig-rashba-exch}(b) and \ref{fig-rashba-exch}(d). This competition between QH gaps of different origins leads to complex and rich quantum phase transitions.

\subsection{Zeeman effect and Rashba spin orbit coupling for $\phi\not=0$}
\label{sect_zeeman_rashba_flux}%

\par
\begin{figure}[t]
\includegraphics{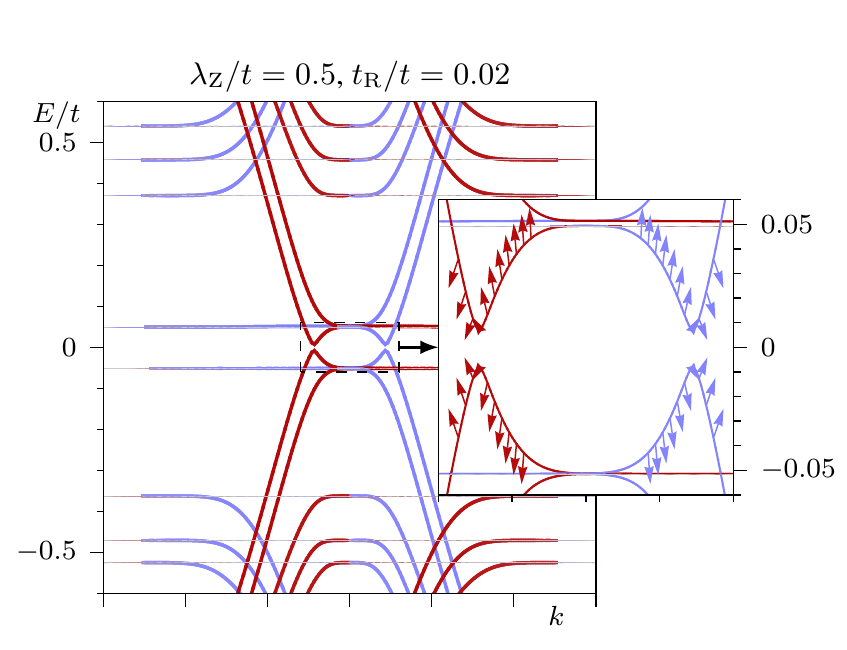}
\caption{\label{fig_dispersions_zeeman_rashba}\coloronline Spectrum for the system with nonzero Zeeman ($\lambdaZ/t=0.5$) and RSO coupling ($\tR/t=0.02$), for $\phi=1/61$. The inset shows a zoom of the region where a gap opens between the edge states at zero energy. For comparison with the spectrum in absence of Rashba coupling, refer to Fig.~\ref{fig_dispersions_zeeman_or_iso}(b).}
\end{figure}

 In the low-energy regime, where the (weak) QSH phase is generated by either ISO coupling or the Zeeman effect, the additional inclusion of the Rashba coupling has several profound effects on the physics. First of all, the RSO coupling tends to decrease the size of the gap (cf.~Ref.~\onlinecite{KaneMele2005PRL95-14}), eventually completely closing it. Secondly, the combination of RSO coupling with ISO coupling and/or Zeeman effect destroys the particle-hole symmetry of the spectrum. Finally, at finite magnetic field, the RSO coupling opens up a gap (i.e., an avoided crossing) between the edge states at their crossing around zero energy. Around this crossing, the spin direction rotates from down to up (or vice versa), as illustrated in Fig.~\ref{fig_dispersions_zeeman_rashba}. The size of this gap is approximately linear in $\phi$ and in $\tR$ (in the limit $\phi\to 0$ and $\tR\to 0$). Thus, the bulk gap is trivial in the sense that no edge states cross it from one bulk band to the other. However, in the limit where the edge-state gap is small, one would still observe a weak QSH state. At finite temperatures, the thermal energy can ``bridge'' the edge-state gap, so that the charge carriers may tunnel from one edge state to another. As the edge-state gap size is increased, the backscattering is enhanced, due to the decrease in the amount of tunneling across the edge-state gap. Here, we expect that the value of the spin Hall conductivity will deviate from its quantized value $\pm2e/4\pi$. The amount of deviation increases when the gap size increases.

\section{Experimental realizations}
\label{sect_experiment}%

In this section, we will discuss the several experimental approaches that allow for the observation of the topological phases and phase transitions discussed in \sects\ref{sect_magnetic_field} and \ref{sect_competition_and_phasetrans}. For each of them, we present the opportunities and challenges, and we discuss the range of parameters that can be probed.

\subsection{Condensed-matter systems}
\label{sect_experiment_condmat}%
\par Condensed-matter systems typically have a lattice spacing in the order of $1\,$\AA, so that one flux quantum per unit cell ($\phi=1$) corresponds to a magnetic field in the order of $10^4$--$10^5\,\mathrm{T}$. Thus, only the low-flux limit is relevant with respect to realistic condensed-matter realizations of our model. The prime example of a two-dimensional system with a honeycomb lattice is graphene, which has received much attention both at the theoretical and experimental levels.\cite{CastroNetoEA2009,Goerbig2011RMP} In the presence of a magnetic field, graphene shows an anomalous quantum Hall effect, where the Landau-level energies\cite{NovoselovEA2005,*ZhangEA2005} depend on the magnetic field as $\sqrt{\phi}$, showing a radically different behavior than in semiconductor heterostructures (e.g., GaAs/AlGaAs).

Besides, the effects of spin-orbit coupling have been extensively studied for graphene, leading to the concept of $Z_2$ topological insulators.\cite{KaneMele2005PRL95-14} However, recent analysis has shown that graphene exhibits only a very weak ISO coupling ($\tI/t\sim10^{-6}$--$10^{-5}$, where $t\sim2.8\,\mathrm{meV}$),\cite{VanGelderenMoraisSmith2010,BoettgerTrickey2007} which unfortunately prevents the observation of the QSH state in this material. Nevertheless, the ISO interactions may be enhanced by putting heavy adatoms on top of the graphene surface,\cite{WeeksEA2011} which would open up a route to build a topological heterojunction.\cite{ShevtsovEA2012} The Zeeman effect is of appreciable order of magnitude, $g\sim2$, which corresponds to gaps of $100\,\mathrm{K}$ at a magnetic field of $30\,\mathrm{T}$.\cite{AbaninEA2006} Values of the Rashba coupling as high as $\tR\sim0.1t$ have been reported for graphene on a Ni(111) substrate.\cite{DedkovEA2008} Adsorption of metal atoms on the graphene surface increases this coupling as well.\cite{DingEA2011} Alternatively, high values of the Rashba coupling (accompanied by the spin re-orientation at the edges) may be induced by curving graphene nanoribbons.\cite{GosalbezMartinezEA2011} Besides, it has been suggested that large Rashba coupling leads to topological phases in bilayer graphene.\cite{QiaoEA2011} In addition, large effective pseudomagnetic fields, greater than $300\,\mathrm{T}$, have been recently realized in strain-induced graphene nanobubbles.\cite{LevyEA2010}  In summary, the observation of the topological phases presented here remains elusive in graphene, but should not be definitively ruled out.

\par The QSH effect has nevertheless been observed in other two-dimensional compounds that have considerably larger values of $\tI$, such as Hg(Cd)Te quantum wells,\cite{KonigEA2007,BernevigEA2006} where values of the order $\tI\sim1$--$10\,\mathrm{meV}$ have been reported.\cite{BruneEA2011,NovikEA2005} Additionally, this material exhibits a strong Zeeman effect as well ($g\sim20$).\cite{KonigEA2008} The spin-filtered QH states have already been observed in these systems at low fields.\cite{ButtnerEA2011} Hg(Cd)Te quantum wells doped with magnetic ions (e.g., Mn) exhibit an exchange interaction leading to the quantum anomalous Hall effect at zero magnetic field,\cite{LiuEA2008PRL101} and to a nonlinear Zeeman effect that causes reentrant behavior of the topological phases.\cite{BeugelingEA2012PRB85}

\subsection{Artificial honeycomb lattices}
\label{sect_experiment_artificial_lattices}%
\par  The experimental limitation of the low flux values can be overcome by studying systems with a larger lattice spacing. Recently, several approaches have led to the engineering of artificial honeycomb lattices. First of all, arrays of quantum dots on a GaAs, arranged in a honeycomb pattern, have been engineered to simulate actual graphene.\cite{DeSimoniEA2010,SinghaEA2011} The lattice spacing in these systems is of the order of $100\,\mathrm{nm}$, so that high flux values ($\phi\sim 1$) are obtained already with small magnetic fields. However, this large lattice constant leads to a small hopping amplitude, which has hampered the observation of the Hofstadter butterfly until now. Patterned superlattices able to probe the Hofstadter regime with usual magnetic field intensities have been realized in the past, but for square geometries.\cite{MelinteEA2004,ZhangEA2006} To probe the physics described here, one needs a superlattice with a smaller lattice constant, grown on a substrate that could lead to the generation of a strong Rashba interaction, for example. In this case, chiral topological states could be probed also away from zero energy, as we discuss in \sect\ref{sect_halfflux}.

Secondly, artificial graphene lattices have been created by manipulating CO molecules with the tip of a scanning tunneling microscope and arranging them in a triangular lattice on a Cu(111) substrate.\cite{GomesEA2012} These molecules repel the electrons on the Cu surface, thus creating a honeycomb array of sites between which the electrons hop. Measurements of the density of states show that the electrons indeed exhibit a Dirac dispersion. The typical lattice constants of these systems are approximately $7$ times larger than in real graphene, thus allowing for flux values $\sim50$ times larger than in graphene, with reasonable magnetic fields. In addition, effective magnetic fields up to $60\,\mathrm{T}$ can be induced by triaxial strain of the artificial lattice. It can be speculated that by using a heavier metal as substrate, it may be possible to engineer such a lattice with strong spin-orbit coupling, leading to the QSH phase.

\subsection{Ultracold atoms in optical lattices}
\label{sect_experiment_optical_lattices}%
\par Although the parameters of the Hamiltonian \eqref{eqn_hamiltonian} can be controlled to some extent in condensed-matter systems, the possibility to engineer the physics discussed here with ultracold atoms in optical lattices is very appealing. The advantage of the ultracold-atom realization of this model is the flexibility to control the parameters separately across a large range, compared to condensed-matter systems where these parameters are generally fixed by the material properties and the sample geometry.\cite{BlochEA2008,BlochEA2012} A second major advantage of ultracold atoms over condensed-matter systems is that the magnetic fields are produced synthetically, i.e., the Peierls phases are not due to a real magnetic field, but induced by external fields acting on atomic internal states.\cite{DalibardEA2011} In this way, large effective magnetic flux can be produced (i.e. $\phi \sim 1$), which allows the system to be studied beyond the low-flux limit.\cite{AidelsburgerEA2011} The spin-orbit couplings can also be synthesized in optical lattices using similar methods.\cite{GoldmanEA2010,*MazzaEA2012,*OsterlohEA2005,*BeriCooper2011} Moreover, in these configurations, the Zeeman coupling can be controlled independently, as it is not related to the synthetic magnetic fields\cite{MakogonEA2012} (see \sect\ref{sect_rashba_exchange}, where we study the extreme situation where the effect of the  Zeeman coupling is investigated in the absence of synthetic magnetic field, in which case this term is called an ``exchange term''). Another important feature offered by optical-lattice setups is the possibility to control the interactions between the particles: by exploiting Feshbach resonances,\cite{BlochEA2008} it is possible to reach the non-interacting regime and from there on, to include interactions in a controlled way to explore the robustness of topological order against interactions. Let us mention that these systems are usually free of disorder, although disorder can also be generated and tailored at will in optical lattices.\cite{SanchezPalenciaLewenstein2010} These outstanding features allowcus to study the TRS-broken QSH phase, namely the survival of the QSH phase in the presence of a TRS-breaking perturbation.\cite{GoldmanEA2012} Indeed, contrarily to condensed-matter experiments, the absence of (magnetic) disorder can stabilize the helical (counterpropagating) edge states even when TRS is broken. Let us also comment on the fact that all the topological phases and edge-state excitations discussed here rely on the existence of robust bulk energy gaps. Therefore, when the temperature of the system approaches the size of the topological gaps of interest, one expects the effects to be strongly affected, and eventually to disappear. For the ultracold temperatures achieved in cold-atom laboratories, the gaps of interest should typically be of the order $\Delta \sim t$, where $t$ is the tunneling amplitude.  Therefore, we anticipate that only the largest topological gaps presented in  our figures should be considered from the cold-atom point of view.

The quantum phase transitions presented in this work could be equally characterized either by a change in the topological invariants associated to the bulk gaps of interest, or by the modification of the edge-state structures. Although cold atoms do not offer the possibility to directly measure transport coefficients, several methods have been proposed to detect the topological invariants related to the quantized conductivities. \cite{UmucalilarEA2008,AlbaEA2011,*ZhaoEA2011} These proposals are based on the fact that non-trivial band structures modify the atomic densities, leading to several signatures that can be directly seen in time-of-flight or density measurements. On the other hand, it is possible to directly probe edge state structures using Bragg spectroscopy techniques.\cite{StanescuEA2009,*LiuEA2010,GoldmanEA2012PRL,BuchholdEA2012} In a recent work,\cite{GoldmanEA2012PRL} it was shown that a state-dependent light probe, focused on the edge of the cloud and transferring angular momentum to the atoms, clearly identifies the presence of chiral edge states in a QH optical lattice.  This method allows us to project the topological edge states on a dark background, making them visible in \emph{in situ} density measurements. In principle, this efficient scheme could be generalized to spin-1/2 systems in order to detect and distinguish between chiral and helical edge structures.

\section{Conclusions}%
\label{sect_conclusion}%
\par We have shown an in-depth analysis of the edge-state structures generated by the combination of the ISO coupling, the RSO coupling, and the Zeeman effect, together with a uniform and perpendicular magnetic field. In \sect\ref{sect_magnetic_field}, we have studied each term individually in order to pinpoint their effect on the quantum Hall states generated by the magnetic field. In the low-field limit, the ISO coupling and Zeeman effect are formally equivalent, and give rise to the weak QSH phase. Away from zero energy, we observe spin-filtered, spin-imbalanced, and ordinary quantum Hall phases. The RSO coupling is different in the sense that it involves spin-flip terms, which causes the spin states to acquire an in-plane component. In addition, the spin direction inside the gap depends on the Fermi energy, and consequently the spin Hall conductivity of the system is ill-defined. In the case of high magnetic fields, we observe Dirac-type regimes away from zero energy. Here, the Zeeman and ISO coupling open weak QSH gaps in a way analogous to the low-flux limit, whereas the RSO coupling opens a trivial gap (unlike the case at low flux, where the spectrum remains gapless). In addition, the topological phase transitions reported in an earlier work\cite{GoldmanEA2012} have been analyzed in more detail here.

\par A combination of the spin-orbit terms and the Zeeman effect leads to subtle competitions, several examples of which have been discussed in \sect\ref{sect_competition_and_phasetrans}. We have shown that, although the Zeeman effect and ISO coupling both give rise to the QSH phase, they do not always reinforce each other, which has been illustrated by the size of the parameter regime of the weak QSH phase in the low-flux limit. At high flux, the Zeeman effect dominates over the ISO coupling, but the effects of the latter are still well visible in the spectrum. Interestingly, a combination of the Rashba term with an exchange coupling leads to nontrivial topological phases at zero magnetic field, away from half-filling. Finally, inclusion of the Rashba coupling in the case where we have a weak QSH gap generated by the Zeeman effect opens a gap at zero energy between the edge states, which provides an illustration of the breakdown of the QSH effect in the absence of TRS and spin conservation.

This work emphasizes the various topological phase transitions and edge-state structures that could be observed in a wide range of materials and quantum emulators. Although we focused our analysis on the honeycomb lattice, where the spin-orbit couplings and the magnetic field configurations are expressed by specific tunneling terms and Peierls phases, we note that our results do not rely on this specific geometry. The topological phases and the phase transitions discussed here are universal, in the sense that they only rely on the inclusion of general ingredients, which can be found or engineered in many different contexts and geometries. Indeed, this universal property can be understood in the low-energy limit, in which various (\emph{a priori}, very different) tight-binding models actually reduce to the same (Dirac-type) equations. In general, we observe that the main ingredients needed to generate such physics are nontrivial gauge fields, which are coupled to the particles of interest (e.g., electrons in materials and atoms in optical lattices). In this work, we have aimed to emphasize the effects produced by each of these ingredients independently: We have investigated the effects of spin splitting, spin mixing, and gauge potentials leading to magnetic or Haldane-type local flux. We hope that our detailed analysis will deepen the understanding of topological insulating phases and that it will motivate further developments in this exciting and rapidly developing field.

\section*{Acknowledgements}
We are grateful to M. O. Goerbig and P. W. Brouwer for useful discussions. WB and CMS acknowledge financial support by the Netherlands Organisation for Scientific Research (NWO). NG thanks the F.R.S-F.N.R.S for financial support.


\appendix*
\section{Geometry and gauge}
\label{sect_geometry_and_gauge}%
Geometrically, the honeycomb lattice is equivalent to a distorted square lattice where each unit cell has the shape of a rhombus. Since there are two inequivalent sets of sites, there are two sites per unit cell, and hence two sublattices, labeled $\subA$ and $\subB$. The sites, labeled as $j$ and $k$ in the Hamiltonian \eqref{eqn_hamiltonian}, may be uniquely identified by two integer labels $n$ and $m$ for the unit cell and $\subA$ or $\subB$ for the sublattice: In the following, we shall write $j=(n,m,\tau)$ (where $n,m\in\mathbb{Z}$ and $\tau\in\{\subA,\subB\}$) to uniquely identify the sites.

\begin{figure}
\includegraphics{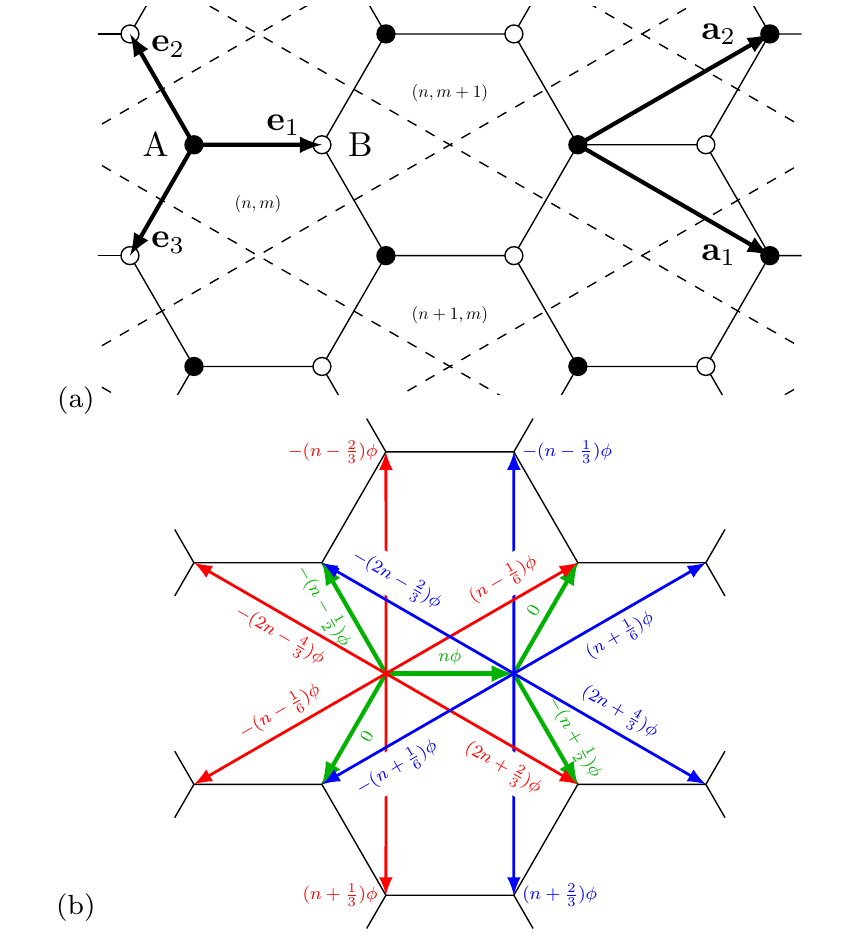}
\caption{\label{fig_lattice}\coloronline (a) The honeycomb lattice. We indicated the NN vectors $\vec{e}_j$ ($j=1,2,3$) and the lattice vectors $\vec{a}_1$ and $\vec{a}_2$. The sites of the sublattices A and B are indicated with solid and open circles, respectively. The dashed lines define boundaries of the unit cells, labeled as $(n,m)$. (b) Hopping phases $\theta_{jk}/2\pi$ for the unit cell $(n,m)$. The phases for the NN hopping are in green. The phases for the NNN hopping are in red and blue for the sublattices $\subA$ and $\subB$, respectively.}
\end{figure}

\par The coordinates of the sites are given as follows. We choose the bond vectors $\vec{e}_1$, $\vec{e}_2$ and $\vec{e}_3$ as displayed in Fig.~\ref{fig_lattice}(a), where their length $a\equiv 1$ is chosen as the unit of length throughout this text. Subsequently, the lattice vectors are given by $\vec{a}_1=\vec{e}_1-\vec{e}_2$ and $\vec{a}_2=\vec{e}_1-\vec{e}_3$, so that the coordinates of the sites are
\begin{equation*}
  \vec{r}_{n,m,\subA}=n\vec{a}_1+m\vec{a}_2,\qquad
  \vec{r}_{n,m,\subB}=n\vec{a}_1+m\vec{a}_2+\vec{e}_1.
\end{equation*}
Furthermore, the reciprocal space is spanned by the two vectors $\vec{b}_1$ and $\vec{b}_2$, defined by $\vec{b}_i\cdot\vec{a}_j=2\pi\delta_{ij}$. We note that the results derived in this paper do not depend on the values of the vectors $\vec{e}_i$; the only effect of a change of the $\vec{e}_i$ is that the lattice vectors and the reciprocal lattice vectors will be different. A convenient choice for the bond vectors is $\vec{e}_1=(0,0)$, $\vec{e}_2=(-1,0)$ and $\vec{e}_3=(0,-1)$, effectively putting both sites of the unit cell $(n,m)$ on the coordinate $(n,m)$.

\par The phases due to the magnetic field, associated with the hopping, are given in terms of the gauge potential $\vec{A}$ as $\theta_{jk}=(e/\hbar)\int_{\vec{r}_k}^{\vec{r}_j}\vec{A}\cdot\vec{d l}$, where the integral is over the line between sites $k$ and $j$. We may employ the gauge freedom for the gauge potential to choose it in such a way that the phase $\theta_{jk}$ for two sites $j=(n,m,\tau)$ and $k=(n',m',\tau')$ (NNs or NNNs) does not depend on the unit cell index $m$. The hopping phases resulting from such a choice are given in Fig.~\ref{fig_lattice}(b) for hopping from the sites of the unit cell $(n,m)$. The reader may check that the phase picked up from hopping around a loop is proportional to the area of the loop. In particular, when one considers one hexagon as the loop, the phase picked up is $2\pi\phi$, which corresponds to an enclosed flux of $\phi$ (in units of the flux quantum). Here, we note that the area of one hexagon is $\tfracs{3\sqrt{3}}{2}$, so that $\phi$ is related to the magnetic field strength $B$ as $\phi=\tfracsb{3\sqrt{3}}{2}\tfracs{eB}{h}$.

\par In addition to the gauge choice, let us furthermore assume that the flux value is a rational number, $\phi=p/q$, where $p$ and $q$ are coprime integers. Then, the phase assigned to each hopping is periodic in the index $n$ with periodicity $q$. In other words, the unit cell of the superlattice (i.e., the lattice together with the hopping phases) is $q\times 1$ unit cells of the original lattice. With this observation, we may invoke Bloch's theorem, which states that we could write the electron wave functions $\varphi_{n,m,\tau}$ as
\begin{equation}\label{eqn_planewave_ansatz}
\begin{aligned}
  \varphi_{n,m,\subA}&=\psi_{n\,\subA}\ee^{\ii\vk\cdot(n\vec{a}_1+m\vec{a}_2)},\\
  \varphi_{n,m,\subB}&=\psi_{n\,\subB}\ee^{\ii\vk\cdot(n\vec{a}_1+m\vec{a}_2+\vec{e}_1)}.
\end{aligned}
\end{equation}
With this ansatz, the Schr\"odinger equation which the fields have to satisfy reduces to a $4q$-component matrix equation: The degrees of freedom are all fields $\Psi_{n}=(\psi_{n\,\subA\,\spinup},\psi_{n\,\subA\,\spindown},\psi_{n\,\subB\,\spinup},\psi_{n\,\subB\,\spindown})$ in the unit cell of the superlattice, so that $n=1,\ldots,q$, and there are two sublattice and two spin components per unit cell of the original lattice. This matrix equation is then merely an eigenvalue equation of a $4q\times4q$ matrix, known as the Harper equation,\cite{Hofstadter1976,ChangYang2004}
\begin{widetext}
\begin{equation}\label{eqn_harper_app}
\frac{E}{t}\begin{pmatrix}\Psi_1\\\Psi_2\\\Psi_3\\\vdots\\\Psi_{q-1}\\\Psi_q\end{pmatrix}
=\begin{pmatrix}
\matD_1&\matR_1&0&\cdots&0&\matR_q^\dagger\\
\matR_1^\dagger&\matD_2&\matR_2&\cdots&0&0\\
0&\matR_2^\dagger&\matD_3&\cdots&0&0\\
\vdots&\vdots&\ddots&\ddots&\ddots&\vdots\\
0&0&0&\ldots& \matD_{q-1}&\matR_{q-1}\\
\matR_q&0&0&\ldots&\matR_{q-1}^\dagger&\matD_q
\end{pmatrix}
\begin{pmatrix}\Psi_1\\\Psi_2\\\Psi_3\\\vdots\\\Psi_{q-1}\\\Psi_q\end{pmatrix}.
\end{equation}
Here, the $4\times4$ matrices $\matD_n=\matD_n^{\mathrm{(NN)}}+\matD_n^{\mathrm{(R)}}+\matD_n^{\mathrm{(I)}}+\matD_n^{\mathrm{(Z)}}$ and $\matR_n=\matR_n^{\mathrm{(NN)}}+\matR_n^{\mathrm{(R)}}+\matR_n^{\mathrm{(I)}}$ encode hopping between unit cells with the same index $n$ and from unit cells with index $n+1$ to $n$. (Note that the Zeeman term is diagonal.) Thus, \eqn\eqref{eqn_harper_app} can be written equivalently as \eqn\eqref{eqn_harper_main}.

\par The $4\times4$ matrices $\matD_n$ and $\matR_n$ may be written in terms of the sum of four contributions, that correspond to each of the terms in Hamiltonian~\eqref{eqn_hamiltonian}. The contributions for the ordinary NN hopping are given by
\begin{equation}
\matD_n^\mathrm{(NN)}
=\begin{pmatrix}
0&\ee^{\ii\vk\cdot\vec{e}_1}\ee^{2\pi\ii n \phi}+\ee^{\ii\vk\cdot\vec{e}_3}\\
\ee^{-\ii\vk\cdot\vec{e}_1}\ee^{-2\pi\ii n \phi}+\ee^{-\ii\vk\cdot\vec{e}_3}&0
\end{pmatrix}
\otimes\idmat_{\mathrm{spin}},
\qquad
\matR_n^\mathrm{(NN)}=
\begin{pmatrix}
0&0\\
\ee^{-\ii\vk\cdot\vec{e}_2}\ee^{-2\pi\ii (n+\frac{1}{2}) \phi}&0
\end{pmatrix}
\otimes\idmat_{\mathrm{spin}}.
\end{equation}
The Zeeman effect is encoded by $\matD_n^\mathrm{(Z)}=(2\pi\lambdaZ\phi/t)\idmat_{\subA\subB}\otimes\sigma_z$ and $\matR_n^\mathrm{(Z)}=0$. The matrices for the ISO coupling are
\begin{align}
\matD_n^\mathrm{(I)}&
=-\ii\frac{\tI}{t}\begin{pmatrix}
-\ee^{-2\pi\ii(n-\frac{1}{6})\phi}\ee^{\ii\vk\cdot\vec{f}_2}+\ee^{2\pi\ii(n-\frac{1}{6})\phi}\ee^{-\ii\vk\cdot\vec{f}_2}&0\\
0&\ee^{-2\pi\ii(n+\frac{1}{6})\phi}\ee^{\ii\vk\cdot\vec{f}_2}-\ee^{2\pi\ii(n+\frac{1}{6})\phi}\ee^{-\ii\vk\cdot\vec{f}_2}
\end{pmatrix}\otimes\sigma_z,\\
\matR_n^\mathrm{(I)}&
=-\ii\frac{\tI}{t}\begin{pmatrix}
-\ee^{2\pi\ii(2n+\frac{2}{3})\phi}\ee^{\ii\vk\cdot\vec{f}_3}+\ee^{2\pi\ii(n+\frac{1}{3})\phi}\ee^{-\ii\vk\cdot\vec{f}_1}&0\\
0&\ee^{2\pi\ii(2n+\frac{4}{3})\phi}\ee^{\ii\vk\cdot\vec{f}_3}-\ee^{2\pi\ii(n+\frac{2}{3})\phi}\ee^{-\ii\vk\cdot\vec{f}_1}
\end{pmatrix}\otimes\sigma_z,
\end{align}
\end{widetext}
where $\vec{f}_1=\vec{e}_2-\vec{e}_3$, $\vec{f}_2=\vec{e}_3-\vec{e}_1$, and $\vec{f}_3=\vec{e}_1-\vec{e}_2$ denote the NNN vectors. Here, the phase factors due to the flux correspond to those in Fig.~\ref{fig_lattice}(b). Due to the fact that the phases involve fractions of $\phi$ which are all multiples of $\tfrac{1}{6}$, the $\phi$-periodicity of the butterfly spectra is $6$, rather than the periodicity of $1$ which is observed in absence of the ISO coupling.
Finally, for the RSO coupling, the hopping matrices are given by
\begin{align}
\matD_n^\mathrm{(R)}&
=\frac{\tR}{t}\begin{pmatrix}
0&0&0&-a_-^\mathrm{(R)}\\
0&0&a_+^\mathrm{(R)}&0\\
0&a_+^{\mathrm{(R)}\,*}&0&0\\
-a_-^{\mathrm{(R)}\,*}&0&0&0
\end{pmatrix},\\
\matR_n^\mathrm{(R)}&
=\frac{\tR}{t}\begin{pmatrix}
0&0&0&0\\
0&0&0&0\\
0&b_+^{\mathrm{(R)}\,*}&0&0\\
b_-^{\mathrm{(R)}\,*}&0&0&0
\end{pmatrix},
\end{align}
where $a_\pm^\mathrm{(R)}=\ee^{2\pi\ii n\phi}\ee^{\ii\vk\cdot\vec{e}_1}+\ee^{\pm2\pi\ii/3}\ee^{\ii\vk\cdot\vec{e}_3}$ and
$b_\pm^\mathrm{(R)}=\ee^{-2\pi\ii (n+\frac{1}{2})\phi}\ee^{\mp2\pi\ii/3}\ee^{\ii\vk\cdot\vec{e}_2}$.

\par The solution of the Harper equation yields the dispersions of the bulk states, which are periodic over the so-called magnetic Brillouin zone, $q$ times smaller than the original Brillouin zone. Effectively, the geometry of the system is a torus, in this case. However, to study edge states, a cylindrical geometry (as displayed in Fig.~\ref{fig_cylinder_edge_states}) is more convenient.  The edge-state spectrum is obtained from the Harper equation similar to \eqn\eqref{eqn_harper_app}, but with a matrix of dimension $4w\times4w$,
where $w$ is the width of the ribbon in unit cells, and with the blocks in the top-right and bottom-left corner replaced by zeros (i.e. $\matR_w\equiv0$). In this case, the dispersion depends on just one component of $\vk$, and its periodicity is the reciprocal lattice vector.

\par To determine states as bulk states or edge states, we derive the density profile $\abs{\Psi_n}^2$ as a function of the unit cell index $n$ (or equivalently, the coordinate component $y$). For convenience, we choose $n=0$ to be the center of the ribbon, so that $n=-\tilde w,\ldots,\tilde w$, where $w=2\tilde w+1$ is the width of the ribbon. The expectation value of this position, $\avg{n}=\sum_n n\abs{\Psi_n}^2$ is used to distinguish edge states and bulk states. Edge states are characterized by a density profile that is sharply peaked at one edge, so that $\avg{n}\approx\pm\tilde w$. On the other hand, if the expectation value $\avg{n}$ is close to zero, then it means that there is a significant contribution of the density away from the edge; such a state is identified as a bulk state. In the edge state spectra of this paper, we have used this expectation value to color the states: Red and blue colors indicate the two opposite edges, while gray is used for the bulk states. In the same way, we use the spin expectation values $\avg{\sigma_i}=\sum_n \Psi_n^\dagger\sigma_i\Psi_n$ ($i=x,y,z$) to gain information about the spin states. We note that this expectation is only reliable in the case that the spin is constant (or almost constant) where the density of the state is concentrated. This may not be the case in the presence of Rashba coupling. In the case when the spin strongly depends on position, the length of the vector $\avg{\vecsigma}$ is less than unity.


%

\end{document}